\documentclass[conference,compsoc]{IEEEtran}
\IEEEoverridecommandlockouts
\usepackage{cite}
\usepackage{amsmath,amssymb,amsfonts}
\usepackage{graphicx}
\usepackage{textcomp}
\usepackage{xcolor}
\usepackage{colortbl}
\usepackage{xcolor}
\usepackage{bm}

\usepackage{lineno}
\usepackage{array}
\newcolumntype{C}[1]{>{\centering\arraybackslash}m{#1}}
\usepackage{CJKutf8}
\usepackage{afterpage}
\usepackage{color}
\definecolor{green}{rgb}{0, 0.5, 0}
\definecolor{orange}{rgb}{0.8, 0.6, 0.2}
\definecolor{orange2}{rgb}{1.0, 0.6, 0.2}
\definecolor{red}{rgb}{1.0, 0.0, 0.0}
\definecolor{teal}{rgb}{0.0, 0.4, 0.4}
\definecolor{purple}{rgb}{0.65,0,0.65}
\definecolor{saffron}{rgb}{0.95,0.75,0.2}
\definecolor{turquoise}{rgb}{0.0,0.5,0.5}
\definecolor{black}{rgb}{0.0, 0.0, 0.0}
\definecolor{gray}{rgb}{0.5, 0.5, 0.5}

\usepackage{threeparttable}

\newcommand{\etal}{\textit{et al}.}

\newcommand{\ie}{\textit{i}.\textit{e}.}
\newcommand{\eg}{\textit{e}.\textit{g}.}

\newcommand{\Tref}[1]{Table~\ref{#1}}
\newcommand{\Eref}[1]{Eq.~(\ref{#1})}
\newcommand{\Fref}[1]{Fig.~\ref{#1}}
\newcommand{\Sref}[1]{Sec.~\ref{#1}}

\definecolor{baselinecolor}{gray}{.9}
\newcommand{\baseline}[1]{\cellcolor{baselinecolor}{#1}}

\usepackage{amssymb}
\usepackage{pifont}
\newcommand{\cmark}{\ding{51}}%
\newcommand{\xmark}{\ding{55}}%

\usepackage{amsmath,amsfonts,bm}

\def\1{\bm{1}}

\def\rvm{{\mathbf{m}}}

\def\rvr{{\mathbf{r}}}
\def\rvs{{\mathbf{s}}}

\def\rvx{{\mathbf{x}}}
\def\rvy{{\mathbf{y}}}
\def\rvz{{\mathbf{z}}}
\def\rvS{{\mathbf{S}}}

\DeclareMathAlphabet{\mathsfit}{\encodingdefault}{\sfdefault}{m}{sl}
\SetMathAlphabet{\mathsfit}{bold}{\encodingdefault}{\sfdefault}{bx}{n}

\def\gN{{\mathcal{N}}}

\usepackage[normalem]{ulem}
\usepackage{multirow}
\usepackage{graphicx}
\usepackage[caption=false]{subfig}
\usepackage[textsize=tiny]{todonotes}
\usepackage{enumitem}
\usepackage{amsmath}
\usepackage{amssymb}
\usepackage[hidelinks]{hyperref}
\usepackage{cleveref}
\usepackage[justification=justified]{caption}
\usepackage{adjustbox}
\usepackage{sidecap}

\newenvironment{packeditemize}{
\begin{list}{$\bullet$}{
\setlength{\labelwidth}{8pt}
\setlength{\itemsep}{0pt}
\setlength{\leftmargin}{\labelwidth}
\addtolength{\leftmargin}{\labelsep}
\setlength{\parindent}{0pt}
\setlength{\listparindent}{\parindent}
\setlength{\parsep}{0pt}
\setlength{\topsep}{3pt}}}{\end{list}}
\hypersetup{
  colorlinks   = true,    
  urlcolor     = blue,    
  linkcolor    = black,    
  citecolor    = red      
}

\def\BibTeX{{\rm B\kern-.05em{\sc i\kern-.025em b}\kern-.08em
    T\kern-.1667em\lower.7ex\hbox{E}\kern-.125emX}}
    
\begin{document}

\title{Catch You Everything Everywhere: Guarding Textual Inversion \\ via Concept Watermarking
}

\author{\IEEEauthorblockN{Weitao Feng$^{1}$, Jiyan He$^{1}$, Jie Zhang$^{2*}$\thanks{* Corresponding author: jie\_zhang@ntu.edu.sg}, Tianwei Zhang$^{2}$, Wenbo Zhou$^{1}$, Weiming Zhang$^{1}$, and Nenghai Yu$^{1}$}
\IEEEauthorblockA{ $^{1}$University of Science and Technology of China \\
$^{2}$Nanyang Technological University \\}}


\maketitle
\begin{abstract}

AIGC (AI-Generated Content) has achieved tremendous success in many applications such as text-to-image tasks, where the model can generate high-quality images with diverse prompts, namely, different descriptions in natural languages. 
More surprisingly, the emerging personalization techniques even succeed in describing unseen concepts with only a few personal images as references, and there have been some commercial platforms for sharing the valuable personalized concept. 
However, such an advanced technique also introduces a severe threat, where malicious users can misuse the target concept to generate highly-realistic illegal images. Therefore, it becomes necessary for the platform to trace malicious users and hold them accountable.

In this paper, we focus on guarding the most popular lightweight personalization model, \ie, Textual Inversion (TI). To achieve it, we propose the novel \textit{concept watermarking}, where watermark information is embedded into the target concept and then extracted from generated images based on the watermarked concept. Specifically, we jointly train a watermark encoder and a watermark decoder with the sampler in the loop.

It shows great resilience to different diffusion sampling processes possibly chosen by malicious users, meanwhile preserving utility for normal use. In practice, the concept owner can upload his concept with different watermarks (\ie, serial numbers) to the platform, and the platform allocates different users with different serial numbers for subsequent tracing and forensics.

Extensive experiments demonstrate that the proposed \textit{concept watermarking} is effective for guarding Textual Inversion meanwhile guaranteeing its utility in terms of both visual fidelity and textual editability. More importantly, our method is robust against different processing distortions, diffusion sampling configurations, and even adaptive attacks. Some ablation studies are also implemented to verify our design. We hope this work can shed some light on guarding personalization models.

\end{abstract}

\section{Introduction}
\label{sec:intro}
\begin{figure}[t]
\centering
\includegraphics[width=0.48\textwidth]{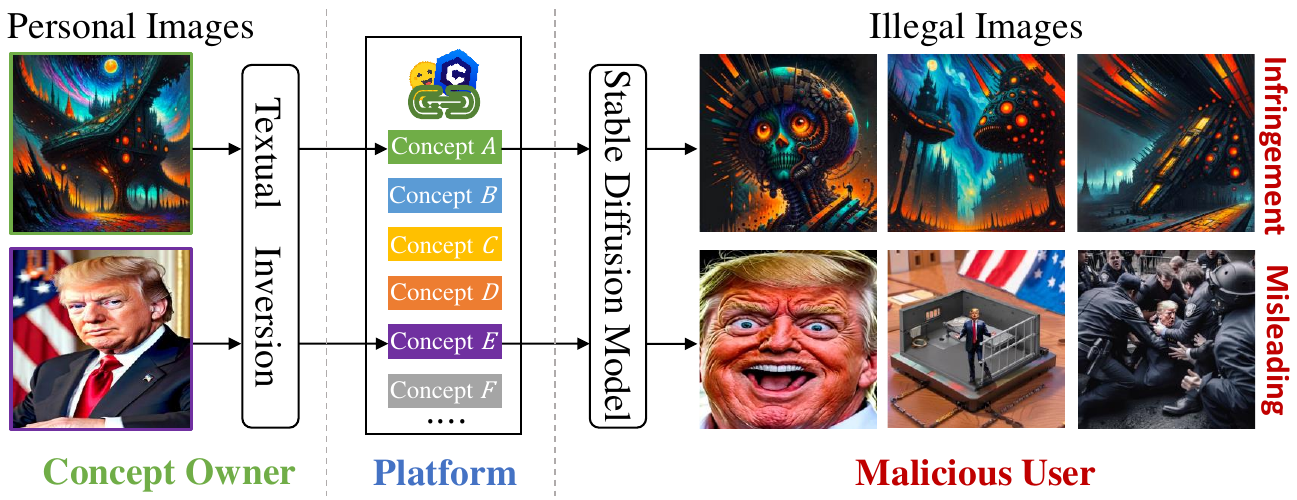}
\caption{The scenario of concept sharing and potential misuses such as infringement and misleading. }
\label{fig:teaser}
\end{figure}

With the rapid progress of generative models, creating high-resolution and highly realistic images from text has become remarkably simple\cite{dhariwal2021diffusion,nichol2022glide,ramesh2022hierarchical,saharia2022photorealistic,rombach2022high}. This has led to a growing demand for generating highly personalized and customized content. With the emergence of various personalization techniques \cite{gal2022image,ruiz2023dreambooth,kumari2023multi,gal2023designing,shi2023instantbooth}, it is feasible to incorporate novel concepts into generative models. 
Sharing platforms for concepts have also emerged such as CivitAI \cite{civitai} and HuggingFace \cite{huggingface}, facilitating the circulation and exchange of a wide range of concepts.
In \Fref{fig:teaser}, we display the scenario of concept sharing, wherein there are three parties: 
1) The concept owner, who owns some personal images and utilizes some personalization techniques such as Textual Inversion \cite{gal2022image} to translate personal images into a concept. 2) The concept-sharing platform provides a chance for concept owner to upload their valuable concepts for sharing or profit. 3) The concept user can download desired concepts and plug them into off-the-shelf Stable Diffusion models for generating concept-related images.

While these advancements demonstrate the rapid progress of generative models, they also raise some severe concerns about their potential misuse by malicious users. 
On the one hand, these concepts may be used to produce images for illegal commercial purposes. As shown in the right top of \Fref{fig:teaser}, the attacker can download concept $A$ representing creative artistic style to generate images with similar style and claim these images as his own hand-drawn artwork, deceiving the public for personal gain \cite{AIartist}.
Because these creative concepts incorporate the creators' expertise and creativity, creators have the right to demand the usage of concepts for certain ranges such as only sharing rather than infringement for commercial purposes.
On the other hand, with the concept of some celebrities, the attacker is able to generate sensitive visuals, or even fake images containing misleading content \cite{t_arrested}, as in the right bottom of \Fref{fig:teaser}.
Therefore, it is necessary to trace the misuse actions and hold the attacker accountable.

\begin{figure}[t]
\centering
\includegraphics[width=0.48\textwidth]{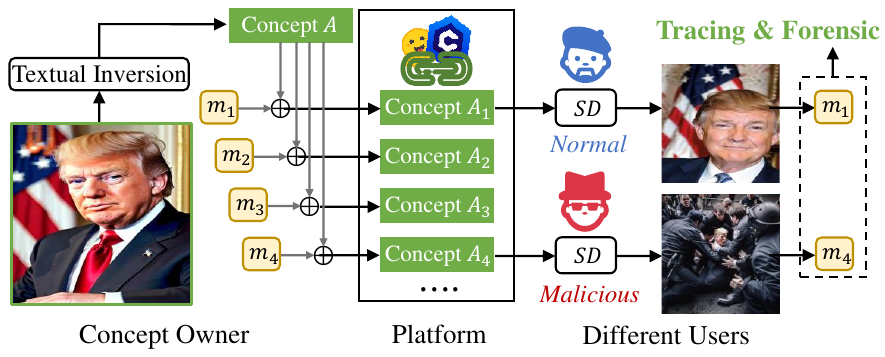}
\caption{The overview of \textit{concept watermarking} for guarding concept sharing. }
\vspace{-1em}
\label{fig:teaser2}
\end{figure}

To address these concerns, we propose the novel \textit{concept watermarking}, which aims to embed watermark information to the concept and extract the watermark from the subsequent generated images based on the watermarked concept. With the proposed \textit{concept watermarking}, we can guard the concept-sharing process with the collaboration between the concept owner and the platform. As shown in \Fref{fig:teaser2}, the concept owner embeds different watermark information into the pristine concept to obtain different concept versions, which are uploaded to the platform. Besides, the decoder related to the concept is also sent to the platform. The platform only needs to allocate different users with different concept versions and builds the relationship between the user ID and the version number. During verification, the watermark can be extracted by the corresponding decoder from the misuse images. Based on the watermark information, the platform can trace the malicious user and provide evidence for forensics and accountability.

To achieve \textit{concept watermarking}, a naive way is to add the same watermark in the personal images before the training of Textual Inversion (as shown in \Fref{fig:teaser}). In fact, adding watermarks to generated content has increasingly become a consensus among large tech companies and governments \cite{openaisec,googlecommitment,Whitehouse}, and some watermarking methods such as DWT-DCT\cite{rahman2013dwt}, DWT-DCT-SVD\cite{rahman2013dwt}, and RivaGAN\cite{zhang2019robust} are also provided by Stable Diffusion official repository \cite{sd} as options. We direct adopt them as our baseline methods but find that all of them totally fail (as displayed in \Tref{tab:performance}). Therefore, it is non-trivial to achieve \textit{concept watermarking}. Here, we point out some  challenges or requirements for \textit{concept watermarking} as follows:

\begin{packeditemize}
    \item  \textit{\textbf{Fidelity.}}  For traditional watermarking \cite{rahman2013dwt,zhu2018hidden,tancik2020stegastamp}, fidelity means preserving visual quality between watermarked images and original images. However, the fidelity of \textit{concept watermarking} means more. Specifically, we shall retain the reconstruction ability and editability of the watermarked concept, which are two common metrics used in personalization tasks \cite{gal2022image,ruiz2023dreambooth}. That is to say, fed with the watermarked concept, text-to-image models can still generate images related to the original concept. Meanwhile, the watermarked concept shall still be able to cooperate with other prompts for image editing.
    
    \item  \textit{\textbf{Effectivness.}} Similar to traditional watermarking, effectiveness requires we successfully extract watermark information from the watermarked images.
    
    \item  \textit{\textbf{Integrity.}} For effective forensics, we shall guarantee to extract the wrong watermark information from the non-watermarked image.

    \item  \textit{\textbf{Robustness.}} Robustness is an important metric to evaluate the utility of watermarking methods in practice. Previous image watermarking methods mainly focus on some digital post-processing distortions (\eg, affine transformations and JPEG) and cross-channel transmission distortions (\eg, printing and scree-shooting). All the above distortions are pixel-space distortions. However, \textit{concept watermarking} faces more severe distortions such as the change of layouts and backgrounds due to the guidance of different prompts. As shown in \Fref{fig:teaser}, the illegal images include the concept of "Trump" but with backgrounds like ``jail" and ``arrest". Moreover, \textit{concept watermarking} shall also resist different diffusion processes including different sampler methods, different steps, model versions, etc. 

\end{packeditemize}

To address the above challenges, we elaborately design our training strategies, loss functions, and model architectures. First, we find that if the watermarked concept is near to the original concept, fidelity can be guaranteed. However, a much closer distance will destroy effectiveness. That is to say, there is a trade-off between fidelity and effectiveness. Therefore, we adopt an incremental training strategy to first achieve acceptable fidelity and then pursue effectiveness. More interestingly, different model architectures hold different performances on fidelity (see \Tref{tab:encoder_arch} and \Fref{fig:t-SNE}), and we adopt the best U-Net as our watermark encoder.
Second, we point out that with an integrity-enhanced decoder, we can successfully achieve integrity so that the model will not falsely extract watermarks from clean images.
\textit{Last but most important},
we introduce a novel sampler-in-the-loop training framework with truncated diffusion sampling (see \Fref{fig:framework} and \Sref{sec:Method}), to obtain the robustness against the diffusion sampling process.
Besides, we find that concept watermarking is inherently robust against most pixel-level distortions. We explain it because the abstract concept is not disrupted by such distortions. Nevertheless, we also adopt widely-used robustness fine-tuning \cite{zhu2018hidden,tancik2020stegastamp} for stronger robustness. 
In addition to the pixel-level distortions, the malicious user may deliberately pre-process the watermarked concept to remove the watermark.
To remedy it, 
we design a contrastive loss to improve the robustness against such pre-processing distortions (See \Tref{tab:Contrastive Loss}).

Extensive experiments demonstrate that the proposed \textit{concept watermarking} can effectively guard Textual Inversion in terms of fidelity, effectiveness, integrity, and robustness. Furthermore, our method can resist some potential adaptive attacks, and many ablation studies are conducted to verify our design. In summary, the primary contributions of our work are concluded as follows:

\begin{itemize}[leftmargin=*]

\item We point out that it is necessary to guard the scenario of concept sharing and propose the novel \textit{concept watermarking} as a feasible solution.

\item To overcome the challenges for \textit{concept watermarking}, we adopt a progressive training strategy to satisfy fidelity and explore an integrity-enhanced decoder for effective forensics. Besides, we find \textit{concept watermarking} is inherently robust against pixel-level distortions. To resist deliberate concept-level distortion, a contrastive loss is also designed.

\item Extensive experiments demonstrate that the proposed method is effective for guarding Textual Inversion comprehensively, even facing adaptive attacks. We also conduct many ablation studies to justify our elaboration.

\end{itemize}

\section{Related Work}

\subsection{Diffusion Models }

Recent advances in generative models have facilitated the emergence of various techniques, including Variational Autoencoders (VAE) \cite{kingma2013auto}, Generative Adversarial Networks (GAN)\cite{GoodfellowPMXWOCB14}, Flow-based Models \cite{rezende2015variational}, and Diffusion Models\cite{sohl2015deep}.
Among them, Diffusion Models (DMs) have achieved state-of-the-art results in image synthesis, which employ non-equilibrium thermodynamics principles to gradually transform a simple prior distribution $q_T$ into a complex one $q_0$ over a preset maximum number of timesteps $T$ in the diffusion process.

\def\imagespace{\mathbb{R}^{H \times W \times 3}}
\def\latentspace{\mathbb{R}^{h \times w \times c}}
\def\interspace{\mathbb{R}^{M \times d_{\tau}}}

The most prevalent variant of diffusion models is Latent Diffusion Model (LDM) \cite{rombach2022high}.
which runs the diffusion process in the latent space, making its training and inference more efficient.
Specifically, LDM utilizes an image encoder  $\mathcal{E}$ to convert an image $\rvx \in \imagespace$ into a latent representation $\rvz$, \ie, $\rvz = \mathcal{E}(\rvx) \in \latentspace$. Simultaneously, an image decoder $\mathcal{D}$ reconstructs the image from the latent representation $\rvz$ in a reverse fashion, \ie, $\rvx = \mathcal{D}(\rvz)$. 
Additionally, a conditional denoising autoencoder, denoted as $\bm{\epsilon}_{\theta}(\rvz_t, t, \rvy)$, generates images given a specific text $\rvy$ as a condition, where $\rvz_t$ signifies the latent representation at a particular time step $t \in \{1,...,T\}$.

During training stage, a squared error loss $\mathcal{L}$ is leveraged to compel LDM to denoise latent representations $\rvz_t := \alpha_t \rvz_0 + \sigma_t \bm{\epsilon}$ as follows:
\begin{align}
\mathcal{L} = \mathbb{E}_{\rvz_0, {\bm{\epsilon}}, t, \rvy}\left[\left\|\bm{\epsilon}_{\theta}\left(\rvz_{t}, t, \rvy\right)-\bm{\epsilon} \right\|_{2}^{2}\right], \label{eq:diffusion-training}
\end{align}
where $\alpha_t$ and $\sigma_t$ represent the parameters of the diffusion process, $\bm{\epsilon}$ is sampled from the Gaussian distribution $\gN(\mathbf{0}, \mathbf{I})$, and the $\bm{\epsilon}_{\theta}(\rvz_t, t, \rvy)$ is implemented as a time- and text-conditional U-Net \cite{ronneberger2015u}. 
Given the trained diffusion models $\bm{\epsilon}_{\theta}(\rvz_t, t, \rvy)$, the sampling procedures, such as DDPM, can be represented by the following equations:
\begin{align}
\rvz_T &\sim \gN(\mathbf{0}, \mathbf{I}), \label{eq:ddpm-infer-1}\\
\rvz_{t-1} = \frac{1}{\sqrt{\alpha_t}} (\rvz_t - & \frac{1-\alpha_t}{\sqrt{1-\bar{\alpha}_t}}\bm{\epsilon}_{\theta}(\rvz_t, t, \rvy)) + \sigma_t \bm{\epsilon},\label{eq:ddpm-infer-2}
\end{align}
while $\bar{\alpha}_t := \Pi_{i=1}^{t}\alpha_i$.  Various sampling methods, including Denoising Diffusion Implicit Models (DDIM) \cite{song2020denoising}, Analytic-DPMS\cite{bao2021analytic}, Euler\cite{karras2022elucidating} and DPM-Solver \cite{lu2022dpm} can utilize the trained model to sample more efficiently and achieve higher-quality results.

In this paper, we conduct experiments on the most popular LDMs, \ie, Stable Diffusion (SD) \cite{sd}. 
As shown in \Fref{fig:teaser}, the user can download an SD from the official website \cite{sd} and enjoy it locally. As mentioned above, there are diverse options for the implementation of SD, which is a challenge for the robustness of \textit{concept watermarking}. In \Sref{sec:ro-to-diff}, we showcase that the proposed method is robust against different diffusion sampling configurations.

\subsection{Textual Inversion}

Personalization can be regarded as an interesting extension to the generic text-to-image task, namely, concept-driven synthesis, which makes it possible to ``personalize'' the image synthesis. In concept-driven synthesis, the model is fed with a set of images sharing the same concept and learns to synthesize images of that concept in novel contexts such as backgrounds and layouts, which are controlled by the text prompts. There are many emerging personalization techniques such as Textual Inversion \cite{gal2022image}, DreamBooth \cite{ruiz2023dreambooth}, and Lora-DreamBooth \cite{hu2021lora}.
Among them, Textual Inversion is the most lightweight personalization technique. Due to its minimal parameter size and excellent editability along with remarkable generation results, Textual Inversion gains widespread popularity. Its small parameter size also contributes to its widespread sharing, making it a mainstream personalization method widely propagated across various online platforms.

Simply speaking, the goal of Textual Inversion is to provide a text representation for the abstract concept of personal images. Then, the representation can be added to the word list of diffusion models, which is responsible for the subsequent generation. 
Specifically, Textual Inversion aims to invert the novel concepts into pseudo-words in the text encoder's embeddings for the plug-in use of off-the-shelf diffusion models like Stable Diffusion. 
Its training objective can be formulated as an optimization problem:
\begin{align}
\rvs^{*}=\underset{\rvs}{\arg \min } \mathbb{E}_{\rvz_0 \sim \mathcal{E}(\rvx), \rvy, \bm{\epsilon} \sim \mathcal{N}(\mathbf{0},\mathbf{I}), t}\left[\left\|\bm{\epsilon}-\bm{\epsilon}_{\theta}\left(\rvz_{t}, t, \bm{\tau}_{\theta'}(\rvy)\right)\right\|_{2}^{2}\right],
\end{align}
where $\rvs^{*}$ is the embedding of the pseudo-word, $\bm{\epsilon}_{\theta}$ stands for the diffusion model, $\rvx$ is one image from a set of target images and $\rvy$ is the condition (\ie, different prompts).

In this paper, we focus on guarding Textual Inversion via \textit{concept watermarking}, which embeds the watermark into the concept learned by Textaul Inversion. When the concept is subsequently fed to Stable Diffusion model by users, we shall be able to extract the embedded watermark from the corresponding generated images, as shown in \Fref{fig:teaser2}.

\subsection{AIGC Watermarking }
There are some attempts on leveraging watermarking techniques to guard AI-Generated Content (AIGC) or the corresponding generative models, \ie, diffusion models.

The former task belongs to generated content detection, which aims to distinguish whether the suspect images are generated or not (\ie, natural images). Different from passive detection, 
watermarking the generated images before releasing them can be seen as proactive detection, where the detection is based on whether the embedded watermark is extracted or not. This strategy is also adopted by many large tech companies and governments \cite{openaisec, googlecommitment, Whitehouse}. For example, Stable Diffusion official repository \cite{sd} provides watermarking options with  some watermarking methods such as DWT-DCT\cite{rahman2013dwt}, DWT-DCT-SVD\cite{rahman2013dwt}, and RivaGAN\cite{zhang2019robust}. We try these methods for \textit{concept watermarking} by watermarking the training images of Textual Inversion. Unfortunately, all these methods fail, and the corresponding results are showcased in \Tref{tab:performance}.

The latter task targets protecting the intellectual property (IP) of the valuable diffusion model itself. There are some attempts that can be mainly divided into three categories: 1) watermarking the entire training data of diffusion models \cite{zhao2023recipe}\cite{cui2023diffusionshield};
2) implanting backdoors into the target diffusion model by specific text-image pairs \cite{zhao2023recipe}\cite{liu2023watermarking};
3) adjust the diffusion sampling process to learn a specific watermarked sampling \cite{wen2023tree}\cite{alemohammad2023self}.
Besides, there are two more related works: a) Fernandez \etal \cite{fernandez2023stable} specifically focus on tracing the users of different LDM versions by allocating them with a corresponding specific watermarked VAE-decoder; b) Ma \etal \cite{ma2023generative} propose to detect whether the personal image is used for training personalization models like Textual Inversion. However, all methods are not feasible for guarding concept sharing (as shown in \Fref{fig:teaser}). 

To the best of our knowledge, we are the first to guard concept sharing and propose the novel \textit{concept watermarking} as an applicable solution. We hope our work can shed some light on the field of AIGC watermarking.

\section{Threat Model}
\vspace{-0.5em}

In this work, we focus on the scenario of concept sharing, as shown in \Fref{fig:teaser}, which involves three parties: (1) a concept owner who possesses the copyright to a particular concept and releases it on the platform for public sharing and commercial profit, (2) the platform that displays different uploaded concepts and provides its users access to downloading concepts they like but stipulates specific terms of legitimate use, and (3) an attacker who downloads the target concept and leverage off-the-shelf Stable Diffusion for malicious use such as re-distributing generated artworks for profit, generating sensitive images about the concept for damage reputations or producing fake and misleading content, all of which could result in substantial harm. In the following part, we specify the ability and goal of each part in detail.

\subsection{Concept Owner's Ability and Goal}
\vspace{-0.5em}

The concept owner owns some valuable personal images such as artworks or portraits, which can be seen as his intellectual properties (IP). As shown in \Fref{fig:teaser2}, he can upload different versions of a concept to the platform and collaborate with it to protect his IP right. Besides, the concept owner knows the potential malicious use conducted by the attacker based on Stable Diffusion, but he has no knowledge of the diffusion configurations set by the attacker.

To guard concept sharing, the owner encodes $n$ pieces of watermark messages $\mathcal{M}=\{\rvm_1,\rvm_2,...,\rvm_n\}$ into the concept $\rvs$, resulting in $n$ watermarked concepts $\rvS^{\mathrm{wm}}=\{\rvs^{\mathrm{wm}}_1,....,\rvs^{\mathrm{wm}}_n\}$. 
The watermarked concepts shall satisfy the following requirements: fidelity, effectiveness, integrity, and robustness, which have been described in detail (See \Sref{sec:intro}) and the corresponding results are provided in \Sref{sec:exp}.

\subsection{Platform's Ability and Goal}
\vspace{-0.5em}

In our threat model, we assume that the platform is trusted and honest, which is practical because the platform wants to keep its reputation. 
The platform cannot distinguish the normal user and malicious user only based on downloading behaviors, but it has the right to stipulate specific terms of use (ToU), such as requiring proper attribution and disallowing commercial and redistribution activities. Besides, the platform is able to store the records of uploading and downloading.

As shown in \Fref{fig:teaser2}, the platform stores various versions of the watermarked concept $\rvS^{\mathrm{wm}}$ uploaded by the concept owner, each one containing different watermarks. When users download a concept, the platform assigns different watermarked concepts to different users and binds the watermarks $\rvm_i$ to the user identification, keeping records for future verification. This enables the ability to trace back potentially ToU-infringing concepts to malicious users.

Upon discovering suspicious image $I^x$, which might violate the owner's terms of use and infringe on their copyright, the platform can extract watermark information from $I^x$ to verify if it was generated using one of the concepts in $\rvS^{\mathrm{wm}}$. For example, if an attacker generates an image $I^x_i$ employing a watermarked concept $\rvs^{\mathrm{wm}}_i$, the platform can extract the watermark $\rvm_i$ from $I^x_i$. Thus, the platform can then identify the user who downloaded the corresponding concept through the records it maintains, before considering further actions such as blocking users or resorting to the law.

\subsection{Attacker's Ability and Goals}
\vspace{-0.5em}

As an attacker, he can easily download the target concept from the platform, and plug it into the local Stable Diffusion for arbitrary image synthesis. Notably, although the attacker has computation resources, he has no access to the training images related to the target concept. If he can collect them, he can directly train his own Textual Inversion, which is out of our scope.

The goal of the attacker is to generate illegal images but shall bypass the platform guarding, namely, removing or disrupting the watermark to interfere with subsequent tracing. In practice, the attacker can utilize different diffusion configurations including prompts, sampler, sampling steps, etc. (see \Sref{sec:ro-to-diff}).
To remove the embedded watermark, the attacker can post-process the generated images such as color jitter, affine transformation, JPEG, etc. (see \Sref{sec:post-p}). Besides, he can also pre-process the watermarked concept by Gaussian noise, rescaling, and smoothing (see \Sref{sec:pre-p}).

We also consider some adaptive attacks, \ie, retraining the concept embedding and conducting forge attacks (see \Sref{sec:aa}).

\begin{figure*}[t]%
\centering
\includegraphics[width=0.95\textwidth]{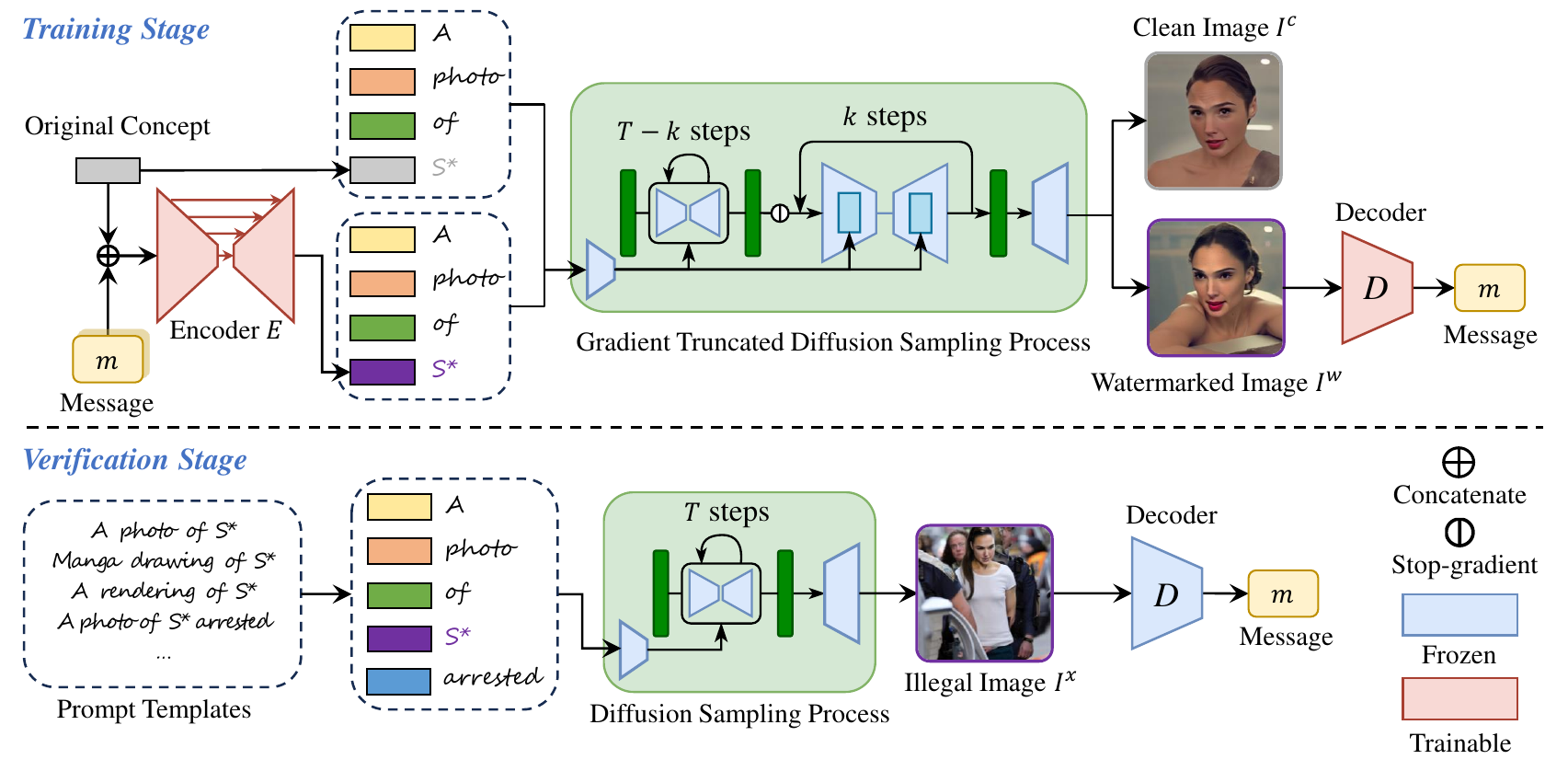}
\caption{The overall framework of the proposed \textit{concept watermarking}. In the training stage, we jointly train the Encoder and Decoder to embed watermarks into Textual Inversion embeddings with sampler in the loop, while ensuring the generation of semantically coherent images $I^c$ and $I^w$. As for verification, we use different prompts as inputs to the diffusion model, and the watermark can be extracted from the generated images $I^x$. }
\label{fig:framework}
\end{figure*}

\section{The Proposed Concept Watermarking} \label{sec:Method}

\def\encoder{{\mathbf{{E}}}}
\def\decoder{{\mathbf{{D}}}}

\subsection{Overview} \label{sec:}

The overall pipeline of our method is illustrated on \Fref{fig:framework}. Above the dashed line is the training stage, wherein the encoder and decoder are jointly trained with the sampler in the loop. Specifically, the random bit string message $\mathbf{m}$ is first replicated and then concatenated with the original concept, serving as the input for the encoder $\encoder$. The encoder targets incorporating the watermark into the embedding, namely, substituting the original embeddings with watermarked embeddings for the corresponding tokenized prompts. The diffusion model accepts encoded prompts with the original concept and watermarked concept as input and generates both clean and watermarked images. For the decoder $\decoder$, it aims to successfully extract the corresponding embedded watermark. 
Below the dashed line is the verification stage. For various prompt inputs from prompt templates, the images generated by our watermarked concept through the diffusion sampling process are required to reveal the corresponding messages contained within the concept.

\subsection{Training Stage}

\subsubsection{Watermark Embedding}

This component is responsible for encoding the input bit string to the original Textual Inversion concept, meanwhile guaranteeing the utility of the watermarked concept.

Our watermark encoder, denoted as $\encoder$, accepts the Textual Inversion concept $\rvs \in \mathbb{R}^{\kappa \times d_\tau}$ and the input message $\mathbf{m} \in \{ 0,1 \}^q$ as its inputs. 
In order to match the size of the Textual Inversion embedding, we replicate the bit message $\mathbf{m}$ by a factor of $\kappa \times d_\tau / q$.
Here, $\kappa$ signifies the number of tokens in the concept, $d_\tau$ represents the representation dimension of the domain-specific encoder, and $q$ indicates the length of the watermarked bit string.
Following this, the repeated message and embedding are merged to form a representation with dimensions $\mathbb{R}^{2 \times \kappa \times d_\tau}$.
The U-Net structure is utilized to enable the effective encoding of the message within the Textual Inversion concept. See more results with different architectures in ablation study \ref{ab:Different Encoder Architecture}.

For training, we incorporate a regularization loss that calculates the $L_2$ distance between the original embedding $\rvs$ and the watermarked embedding $\rvs^{\rvm}=\encoder(\rvs,\rvm)$:
\begin{equation}
L_{\mathrm{Reg}} = \lVert \rvs - \encoder(\rvs,\rvm) \rVert_2^2,
\end{equation}
whose purpose is to facilitate rapid convergence of the encoder output to the original embedding during the early stages of training, enabling a favorable starting point and significantly accelerating the training process.

Additionally, to improve the latent-level robustness, we devised the contrastive loss, which aims to increase the distinction between Textual Inversion embeddings with different watermarks. It is calculated similarly to contrastive learning\cite{chen2020simple}, but within only negative pairs. The similarity is defined as the cosine values of two vectors, which are obtained by subtracting the original Textual Inversion embedding from the watermarked Textual Inversion embeddings. The formulation is as follows:
\begin{align}
\rvr_i &= \frac{\encoder(\rvs,\rvm_i) - \rvs}{\lVert \encoder(\rvs,\rvm_i) - \rvs \rVert_2}, \quad i= \{ 1,...,N\} ,\\
L_{\mathrm{Cst}} &= \frac{1}{N(N-1)} \sum_{i,j=0,i\ne j}^{N} <\rvr_i,\rvr_j> , 
\end{align}
where $N$ denotes the batch size and $\rvm_i$ refers to the $i$-th watermark message employed by the encoder. 
Incorporating this loss enhances the robustness of the embedding against attacks that directly manipulate the embedding. By lessening the similarity between embeddings containing different messages, it becomes difficult to tamper with the embedding and cause misjudgments by the decoder. This is demonstrated in the ablation study \ref{Impact of Contrastive Loss}.

\subsubsection{Gradient Truncated Diffusion Sampling}

We select Stable Diffusion \cite{rombach2022high} as our generative model. During the training process, we freeze the parameters of the diffusion model $\bm{\epsilon}_{\theta}(\rvz_t, t, \rvy)$.

To make the watermark training feasible with the frozen $\bm{\epsilon}_{\theta}(\rvz_t, t, \rvy)$, in which only a part of $\rvy$ is influenced by the watermark encoder $\encoder$, we maintain differentiability throughout the sampling process (e.g., DDPM in \Eref{eq:ddpm-infer-1} and \Eref{eq:ddpm-infer-2}). Simultaneously, we strike a balance by propagating gradients exclusively through the final $k$ sampling steps to ensure memory and computational efficiency, as well as effectiveness. The default setting for $k$ is 3, and this is termed Gradient Truncated Diffusion Sampling (see \Fref{fig:framework}).
This strategy is motivated by the previous investigation on diffusion models \cite{balaji2022ediffi}: in the early stage of the diffusion model sampling, it heavily relied on text prompts to generate text-aligned content. However, in the later stage, text conditioning was almost completely disregarded, and the focus shifted toward producing high-fidelity visual outputs.

During the training process, we let the diffusion model simultaneously forward multiple embeddings in a single mini-batch, including an original embedding and several embeddings with different randomly generated bit strings as watermark messages.  It is essential that the generated images from the watermarked embeddings reflect the same concept as the original embedding. To guarantee it, we employed the CLIP \cite{radford2021learning} to calculate the cosine-similarity loss, which constrains the similarity between the watermarked image and the original image:
\begin{align}
L_{\mathrm{CLIP}} = - \frac{1}{N} \sum_{i=1}^{N} \frac{<\encoder_{\mathrm{CLIP}}(I^c),\encoder_{\mathrm{CLIP}}(I^w_i)>}{\lVert \encoder_{\mathrm{CLIP}}(I^c) \rVert_2 \cdot \lVert \encoder_{\mathrm{CLIP}}(I^w_i) \rVert_2} ,
\end{align}
where $I^c$ and $I^w$ denote the clean image and watermarked image, $\encoder_{\mathrm{CLIP}}$ denotes the CLIP vision encoder.

\subsubsection{Watermark Extraction}

In the watermark extraction phase, the decoder $\decoder$ receives watermarked images generated by the diffusion model and aims to extract the embedded watermark from these images. We employ EfficientNet-B3\cite{tan2019efficientnet} pretrained on ImageNet\cite{deng2009imagenet} as our watermark extraction network. 
We define $q$ as the number of the message-bits encoded into the watermark and turn the output dimension of $\decoder$ into size $2q$. Then, we compute the Binary Cross Entropy (BCE) between the output logits $l_{j}$ of decoder $\decoder$ and the pre-defined watermark message $\rvm$:
\begin{align}
l_{j} &= \frac{\exp({\decoder(I^w)^{(2j-1)}})}{\exp({\decoder(I^w)^{(2j-1)}})+\exp({\decoder(I^w)^{(2j)}})} , \\
\mathcal{L}_{\mathrm{Msg}} & = - \frac{1}{q} \sum_{j = 1}^{q} \rvm^{(j)} \log \left(l_{j}\right)+\left(1-\rvm^{(j)}\right) \log \left(1 - l_j \right) .
\end{align}
During the inference stage, we only need to compare the values of $l_{2j-1}$ and $l_{2j}$ to determine whether the bit $\rvm^{(j)}$ at index $j$ should be 0 or 1.

\subsubsection{Training Details} 

Here, we introduce some training details including training objectives and an additional training strategy, \ie, robustness fine-tuning.

\noindent\textit{\underline{Training Objectives.}}
In the training phase, we design the progressive loss function to tackle the challenges in concept watermarking. In the early iterations of training, we added a strong $L_{\mathrm{Reg}}$ term to ensure that the output of the encoder quickly fits the Textual Inversion embedding, which contains the original concept. After that, we removed the $L_{\mathrm{Reg}}$ term and only kept $L_{\mathrm{CLIP}}$, $L_{\mathrm{Msg}}$ and $L_{\mathrm{Cst}}$. Specifically, we maintained a counter $u$ that decrease 1 when $L_{\mathrm{Reg}}$ is below a specified threshold $h$. When the counter becomes less than 0, it indicates that the fitting is good, and the encoder's output is at a favorable starting point. At this point, we removed the $L_{\mathrm{Reg}}$ term. Experimentally, this training strategy significantly accelerates the fitting speed and produces better results.
In general, our loss function can be formulated as follows:
\begin{align}
L_{\mathrm{Total}} = \left\{\begin{array}{ll}
L_{\mathrm{CLIP}} + L_{\mathrm{Msg}} + \lambda L_{\mathrm{Reg}} + \mu L_{\mathrm{Cst}} & \text { if } u > 0 \\
L_{\mathrm{CLIP}} + L_{\mathrm{Msg}} + \mu L_{\mathrm{Cst}} & \text { else },
\end{array}\right.
\end{align}
where $\lambda, \mu$ are hyper-parameters to weight each term.

\noindent\textit{\underline{Robustness Fine-tuning.}}
The proposed method can inherently resist many pixel-level distortions, as our watermark exists on the concept level. Nevertheless, we further introduce the popular robustness fine-tuning strategy for robustness gain. Briefly,  
prior to inputting generated images into the decoder, we append a noise layer to introduce various types of noise to them. 
During the robustness fine-tuning, we freeze the encoder part and truncated the gradient backpropagation between the distortion layer and the decoder. As there's no need to propagate to the noise layer, we can introduce non-differentiable noise into the decoder's input, enabling us to incorporate non-differentiable post-processing operations seamlessly into our framework. The corresponding results are provided in \Tref{tab: Robustness fine-tuning}.

\subsection{Verification Stage}

In the verification stage, under the setting of diverse prompts, we employ a new prompt template that differs from the prompts used during the training phase. Here, we use the watermarked embeddings $\rvs^{\mathrm{wm}}$ in combination with the prompts from the prompt template. Through a diffusion sampling process, we generate images and subsequently use the decoder $\decoder$ to verify whether the extracted message $\rvm$ matches the watermarked message.

\section{Experiments}
\label{sec:exp}
\subsection{Experiment Settings}
\noindent \textbf{Dataset.}
We collected and curated various types of concepts from online sources as our test subjects. Among them, 5 concepts are specific individuals, 2 concepts are related to art styles, and 3 concepts are about certain objects, including vehicles and rare items. The tested concepts cover mainstream Textual Inversion concept types to a large extent.

For the prompt to generate images, we apply basic prompts such as (``A photo of S*'') as references and employed the GPT-3.5 model\cite{ouyang2022training} to generate more prompt templates that encompassed a wider range of scenes. This approach can involve more prompts in the training process, allowing the watermark to be preserved better in the generated images when the concept is applied to different prompts. More details of prompt generation can be found in Appendix \ref{app:Prompt generation}.

\noindent \textbf{Baseline.}
A straightforward idea is to investigate whether existing image watermarks can be learned during the Textual Inversion training process and subsequently extracted from the generated images. To examine this, we conducted experiments using DWT-DCT\cite{rahman2013dwt}, DWT-DCT-SVD\cite{rahman2013dwt}, and RivaGAN\cite{zhang2019robust} as representative image watermarks due to their good extraction rates and robustness. They are also adopted as the official method of adding watermarks to the generated images in Stable Diffusion. We added these watermarks to all images in the training dataset and then proceeded with the Textual Inversion training. Upon completing the training, we use Stable Diffusion to sample images and attempted to extract the watermarks from them.

It is worth noting that for the baseline implementation, it is necessary to retrain a new Textual Inversion using the original training data. Therefore, we conducted experiments solely on the concepts trained by ourselves rather than the downloaded concepts.

\noindent \textbf{Evaluation Metrics.}
To evaluate our proposed concept watermarking method, we adopt three metrics as follows:
\begin{packeditemize}

    \item  \textit{\underline{Watermark extraction ability.}} To evaluate the correctness of the extracted watermark, we first adopt \textbf{Bit Error Rate (BER)}, which is calculated between the pre-defined watermark bits and the extracted watermark bits. In the subsequent experiments, we only showcase BER, \ie, the average BER calculated among 256 test images. For each test image, the watermark is regarded as successfully
    extracted if it is totally the same as the label. Based on it, the \textbf{Success Rate (SR)} is further defined as the ratio of watermarked images whose hidden watermark is successfully extracted (\textbf{BER}=0) in the test image set.

    \item  \textit{\underline{Visual fidelity.}} We require high visual similarity of images generated by watermarked concept compared with that generated by the corresponding pristine concept.
    To evaluate it, we follow \cite{gal2022image} to adopt \textbf{Image-alignment (I-A)}, which is defined as the cosine similarity of CLIP image embedding between two images. Specifically, we calculate \textbf{I-A} between the 64 generated watermarked images and 64 generated clean images based on basic prompts.

    \item  \textit{\underline{Textual editability.}}
    Finally, we want to preserve the ability to modify the generated contents using prompts. To this end, we produce a set of images using prompts of varying scenes. These range from simplest descriptions (“A photo of S*”), to style changes for non-style concepts (“A colorful graffiti of S*”), and a compositional prompt (“A photo of S* playing guitar in the forest”).
    We measure the alignment between the generated images and the given prompts, ensuring that the alignment does not decrease with the watermark add-on. For this, we use the \textbf{Text-Alignment (T-A)}, which is used in custom diffusion\cite{kumari2023multi} and many other works. It is calculated by given prompts and corresponding images, computing text-image similarity in CLIP feature space.

\end{packeditemize}

\noindent \textbf{Implementation Details.}
For Textual Inversion embedding, it is important to note that different token lengths can exist, where larger token lengths have greater capacity and can embed more bits of information. In our experiment setting, we were able to encode 8 bits of information into a single token-length embedding. We set the gradient backward steps $k$ to 3, $\lambda$ value to 10 and the $\mu$ value to $1 \times 10^{-2}$. The counter $u$ was set to 200, and threshold $h$ is set to $1 \times 10 ^{-2}$. We used the Adam\cite{AdamKingmaB14} optimizer with a batch size of 8. The initial learning rate was set to $1 \times 10^{-4}$ and we use 2 gradient accumulation steps. We chose Stable Diffusion v1.5 as our base model due to its widespread usage and we adopt fp16 for it. All our experiments can be conducted on a single A6000 GPU.

During the training process, we utilized DDIM \cite{song2020denoising} as our sampling method, which, with a reduced number of steps, can generate high-quality images. Although we exclusively used DDIM during training, the watermark exhibited robustness across various samplers after training completion. Further details can be found in the \Sref{Robustness}.

During the robustness fine-tuning stage, as we only need to fine-tune the decoder, the required GPU memory significantly decreases. We set the batch size to 12 and perform fine-tuning for each concept, limiting the steps to a maximum of 8,000. For the rest of the parameter settings, we keep them consistent with the training stage.

\begin{figure*}[t]%
\centering
\includegraphics[width=1.0\textwidth]{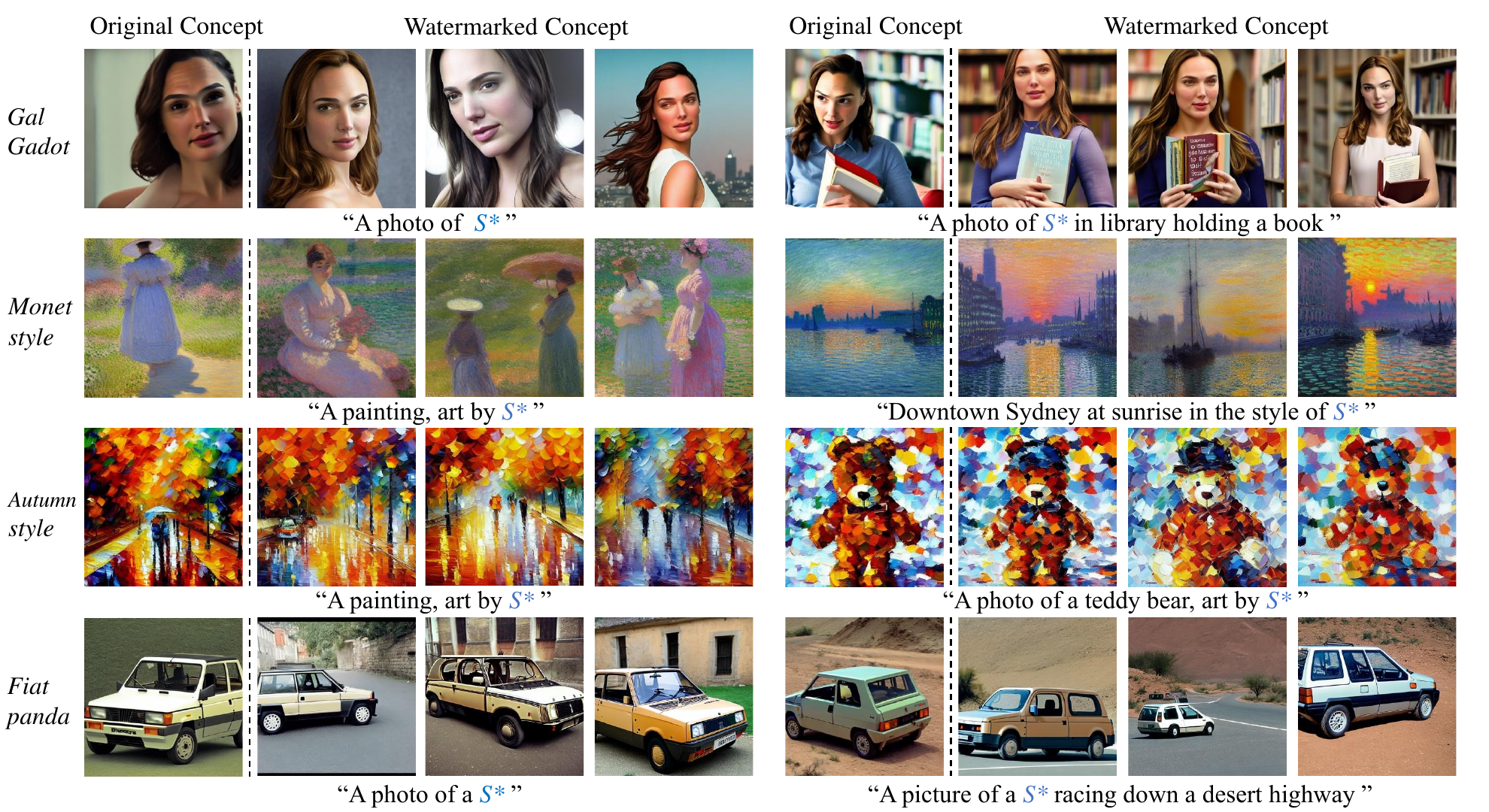}
\caption{Comparison between images generated by the original concept and the watermarked concept under basic prompts and diverse prompts. The image comprises three categories: person, style, and object. \textbf{Left:} The results obtained from basic prompts demonstrate excellent consistency in terms of content semantics. \textbf{Right:} The results from diverse prompts showcase that the concepts with added watermarks maintain the same level of editability as the original concepts.}
\label{fig:result_showcase}
\end{figure*}

\subsection{Effectivness \& Fidelity}

\subsubsection{Effectivness}
\Tref{tab:performance} shows the comparison between our method and the baseline approaches in terms of the BER and Success Rate. The watermark information added by the baseline methods is largely lost during the training process, making it challenging to extract. In contrast, our method excels at effectively extracting the watermark, showcasing its superior performance in guarding concept sharing.

\begin{table}[]
\centering
\caption{Comparison with some baseline methods.}
\setlength{\tabcolsep}{6.5pt}
\begin{tabular}{lcccc}
\hline
\textbf{Method} &  \textbf{BER}(\%)$\downarrow$ & \textbf{SR}(\%)$\uparrow$ & \textbf{T-A}$\uparrow$ & \textbf{I-A}$\uparrow$ \\ \hline
Original                   & -            & -                     & 25.52                   & 81.45                    \\ \hline
TI+DWT-DCT\cite{rahman2013dwt}               & 50.0        & 0.0 (\xmark)        & 25.32                   & 79.88                    \\
TI+DWT-DCT-SVD\cite{rahman2013dwt}           & 50.1        & 0.0 (\xmark)        & 25.93                   & 80.75                    \\
TI+RivaGAN\cite{zhang2019robust}              & 52.1        & 0.0 (\xmark)        & 26.05                   & 81.82                    \\ \hline
Ours                       & 0.3        & 97.4 (\cmark)        & 24.78                   & 79.73                    \\ \hline
\end{tabular}
\label{tab:performance}
\end{table}

\subsubsection{Reconstruction Ability}

\Fref{fig:result_showcase} illustrates qualitative examples of the images generated by watermarked Textual Inversion embeddings and the original Textual Inversion embeddings. They are identical at the concept level, allowing the creators to release the watermarked embeddings instead of the original ones, resulting in images generated from these embeddings containing concept-level watermarks.
In \Tref{tab:performance}, we present the quantitative results, where the calculation for Image-alignment of the original concept was computed by generating 128 images independently using the original concept with the basic prompt and subsequently calculating pair-wise clip similarity between the first 64 and the last 64 generated images. It sets an upper bound for this metric since the ultimate goal is to be as similar to the original concept as possible. Our method yields results that are very close to the upper bound.

\subsubsection{Textual Editability}

Moreover, concerning Text-alignment, our approach achieves favorable scores, indicating that the impact of our watermarks on editability is minor.
The good results obtained by the baseline methods in Text-alignment and Image Alignment can be attributed to the fact that the added watermarks did not significantly impact the training process. As a result, the generated Textual Inversion embeddings closely resemble the original Textual Inversion embeddings, leading to a very similar performance in the evaluation metrics.

\subsection{Integrity}

\begin{figure*}[]
\centering
\includegraphics[width=1.0\textwidth]{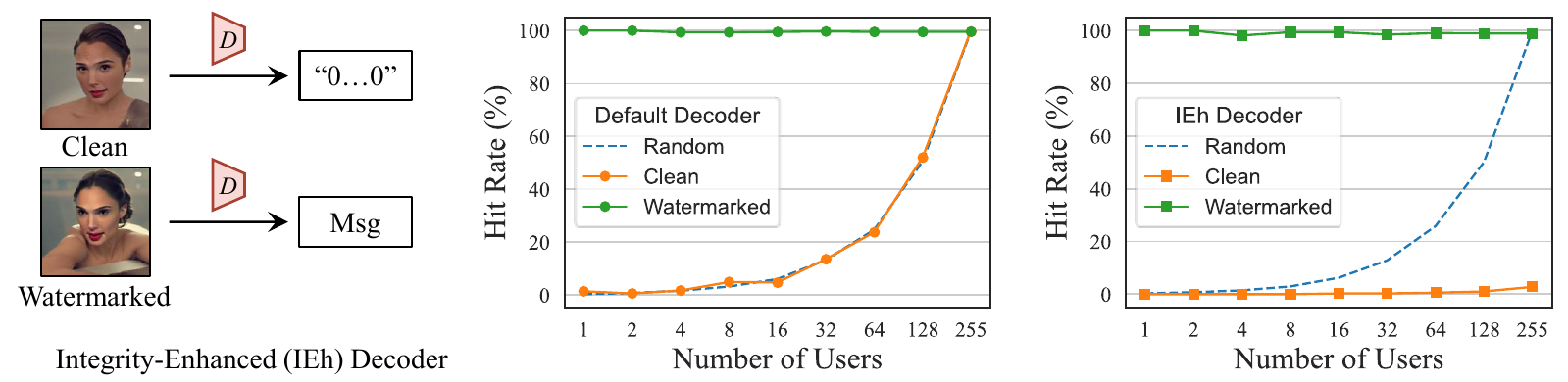}
\caption{
Comparison of Hit Rates for different settings. We include clean images and watermarked images with the default decoder and the Integrity-Enhanced (IEh) decoder. The left figure illustrates the new mapping relationship of our IEh decoder. The middle and right figures show the Hit Rate values when using the default decoder and IEh decoder, respectively, for allocating different watermarked embeddings to different users.
}
\label{fig:integrity_showcase}
\end{figure*}

The integrity of the watermark in our concept-sharing scenarios refers to the accurate identification of specific watermarks from images. This ensures that watermarks are not falsely identified in images that were not encoded using our message encoder, such as those generated with the original concept, other concepts, or other encoders.

To evaluate the integrity of our method, we assume that the platform will distribute various quantities of the same concept with different added watermarks to users. For evaluation, we randomly select $v$ watermarked embeddings and treat them as distributed embeddings. We record their corresponding bit strings. Subsequently, we sample 1000 clean images and 1000 watermarked images using $v$ ``distributed'' embeddings, then attempt to extract information using the decoder. A hit is recorded when the extracted bit string matches any of the $v$ selected bit strings.  For clean images, a lower Hit Rate indicates better performance, and vice versa. We define ``Random'' as the decoded message that is entirely random, calculated by generating 1000 random bit strings. Notably, as our results demonstrate on clean images, the Hit Rate is close to the random scenario.

As depicted in \Fref{fig:integrity_showcase}, when the default decoder is applied, the results for clean images closely resemble random decoding, which is reasonable but not ideal. Thus, we propose an Integrity-enhanced (IEh) version of our decoder, which efficiently prevents the decoder from falsely decoding watermarks from clean images. In this IEh version, we fine-tune the decoder by constraining clean images to be fixed to a specific bit string (\eg, ``00...0"), and the embedding corresponding to this bit string is removed from potentially distributed embeddings. As presented in the left plot of \Fref{fig:integrity_showcase}, we set the special bit string to all zeros. During platform distributing embeddings, at most $2^q-1$ embeddings can be distributed, and the output for clean images is assumed to be the all-zero bit string. With this setup, the falsely identified hit rate was reduced to 2.83\% when releasing 255 concepts, as depicted by the orange line on the right plot of  \Fref{fig:integrity_showcase}.

We also tested if this fine-tuning would impact the decoder's performance on watermarked images. As shown by the green line on the right plot, the performance exhibited minimal change, only a tiny decrease.

\begin{figure*}[t]
    \centering
    \includegraphics[width=\textwidth]{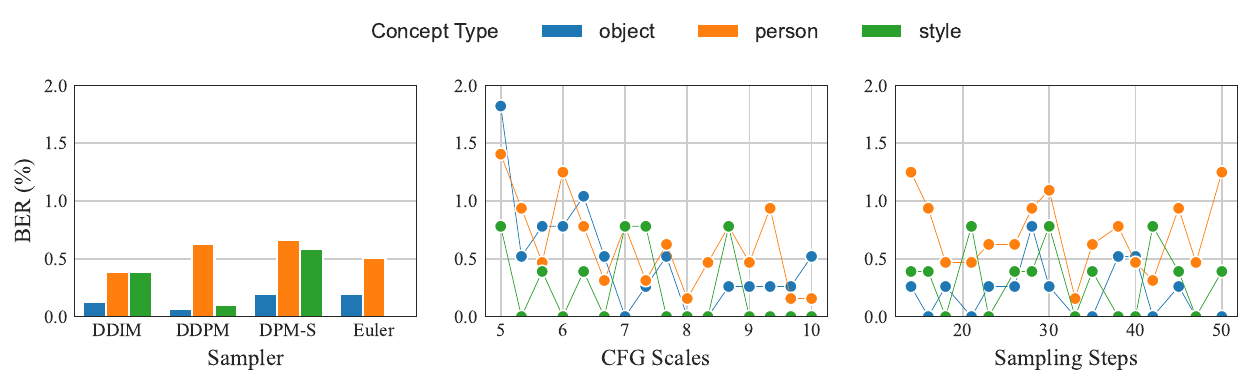}
    \caption{Watermarked concepts demonstrate robustness under diverse configurations, including distinct samplers, CFG scales, and sampling steps. Each color signifies a concept type. The results are quantified using Bit Error Rate (BER).}
    \label{fig:boxplot}
\end{figure*}

\begin{table}[t]
\centering
\caption{Robustness against different diffusion configurations. For each setting, we showcase the average results. The gray cell denotes default setting.}
\setlength{\tabcolsep}{7.0pt}
\begin{tabular}{ccccc}
\hline
\multicolumn{2}{c}{\textbf{Configurations}} & \textbf{BER(\%)$\downarrow$} & \textbf{SR(\%)$\uparrow$} & \textbf{I-A$\uparrow$} \\ \hline
\multicolumn{2}{c}{Default}                   & 0.3                    & 97.4                           & 79.73                  \\ \hline
\multicolumn{2}{c}{Diverse Prompts}           & 3.6                    & 83.7                           & -                      \\ \hline
\multirow{4}{*}{Sampler}          & \baseline DDIM      & 0.3                    & 97.4                           & 79.73                  \\
                                  & DDPM      & 0.3                    & 97.5                           & 79.07                  \\
                                  & DPM-S     & 0.5                    & 96.4                           & 77.63                  \\
                                  & Euler     & 0.3                    & 97.7                           & 79.05                  \\ \hline
\multirow{4}{*}{Sampling Steps}   & 14  & 0.8                    & 94.4                           & 79.81                  \\
                                  &\baseline 25  & 0.3                    & 97.4                           & 79.73                  \\
                                  & 38  & 0.5                    & 95.6                           & 78.58                  \\
                                  & 50  & 0.7                    & 95.0                           & 78.56                  \\ \hline
\multirow{3}{*}{CFG Scales}       & 5.0   & 0.9                    & 93.3                           & 79.87                  \\
                                  &\baseline 7.5   & 0.3                    & 97.4                           & 79.73                  \\
                                  & 10.0  & 0.2                    & 98.1                           & 78.16                  \\ \hline
SD Version                        & v1.5 $\rightarrow$ v1.4  & 1.7                    & 87.5                           & 79.49                  \\ \hline
\end{tabular}
\label{tab:differentdiffconfigs}
\end{table}

\subsection{Robustness} \label{Robustness}
In this section, we evaluate the robustness of the proposed method in a comprehensive way, namely, evaluating robustness against different diffusion sampling configurations, post-processing on generated images, and pre-processing on watermarked concepts.

\subsubsection{Robustness Against Diffusion  Sampling Configurations}
\label{sec:ro-to-diff}

In this section, we explore various settings for the diffusion sampling process. We consider that users can utilize different prompts, samplers, sampling steps, Classifier-Free Guidance (CFG) scales, and different SD versions.

\noindent \textit{\underline{Different Prompts.}}
To test different types of concepts, we generate distinct prompts for each category. Broadly, our concepts fall into three categories: person, style, and object. Since these categories exhibit significant differences and we aim to ensure prompt diversity. Thus for each category, we provide an instruction template and employ GPT-3.5 to generate specific prompts. We generate 64 prompts for each category and sampled 256 images for each concept under different watermark messages. To be specific, we randomly selected 4 different bit strings and get the corresponding watermarked embeddings and sample 64 images for each. More details can be found in Appendix \ref{app:Prompt generation}
As shown in \Tref{tab:differentdiffconfigs}
despite the high diversity of the prompts, our approach still manages to maintain a good extraction rate.

\noindent \textit{\underline{Different Samplers.}}
In the denoising process, various types of samplers can be employed. In our study, we opted for DDIM\cite{song2020denoising}, DDPM\cite{ho2020denoising}, DPM-solver (DPM-S)\cite{lu2022dpm}, and Euler Scheduler\cite{karras2022elucidating}. Among these, DPM-solver and Euler Scheduler are currently the most commonly used and widely adopted samplers. For each sampler, we generated 64 images for our evaluation. We found that changing the sampler has minimal impact to the watermark extraction (see \Tref{tab:differentdiffconfigs}).

\noindent \textit{\underline{Different Sampling Steps.}}
Users have the flexibility to use different sampling steps for content generation. Thus, we evaluated watermark extraction performance under various sampling step settings. We observed that when the sampling steps are fewer than 14, the DDPM sampler fails to produce high-quality results. Consequently, we chose 16 distinct sampling step values that were almost uniformly spread between 14 and 50. For each step value, we generated 4 different images, resulting in a total of 64 images, computed the BER and Success Rate of watermark extraction, and Image-alignment for visual fidelity.
In our study, the watermark demonstrated no signs of decreased extraction rate with changes in the sampling steps (see \Tref{tab:differentdiffconfigs}).

\noindent \textit{\underline{Different CFG Scales.}}
Classifier-Free Guidance (CFG) \cite{ho2022classifier} is a sampling method that offers greater flexibility compared to classifier guidance and has been proven to achieve better generation quality. It has been extensively applied in various diffusion-based generative models. We explored the scenario where users employ different CFG scales. We selected 16 different CFG scale values evenly spaced between 5.0 and 10.0. Similarly, for each CFG scale value, we generated 4 different images, resulting in a total of 64 images, and computed the BER and Success Rate of watermark extraction, Image-alignment for visual fidelity. We observed that although there is a certain tendency for the watermark extraction and CFG scales, our extraction ability still remains at a relatively high level. Furthermore, excessively low CFG scales may lead to unusable results.

\noindent \textit{\underline{Different Versions of Stable Diffusion.}}
The users have the flexibility to choose different versions of Stable Diffusion, such as Stable Diffusion v1.4 or other privately fine-tuned models. To assess our method's ability to withstand such variations, we tested the scenario where the model used in the sampling process was switched to Stable Diffusion v1.4. As shown in  \Tref{tab:differentdiffconfigs}, our method still maintains a good BER. However, it is important to note that for models fine-tuned on datasets with strong personal style or other specific scenarios which have stronger biases, the watermark extraction performance may be impacted more.

We have recorded the BER results under different settings in \Fref{fig:boxplot}. Overall, our method achieves consistently low BER values. We observed that concepts belonging to the ``object'', ``person'', and ``style'' categories show different levels of difficulty, with ``person'' concepts having a higher BER compared to the other two categories. Among all the samplers we tested, DPM-S showed the worst watermark extraction performance, possibly due to larger differences between DPM-S and DDIM compared to DDPM and Euler with DDIM. Regarding the sampling steps, we did not find significant differences between the 14 steps and 50 steps. However, higher CFG scales lead to better extraction results. This is because higher CFG scales increase the influence of prompts, allowing our added watermark to be better expressed.

\subsubsection{Robustness Against Different Post-processing on Generated Images}
\label{sec:post-p}

\begin{table}[t]
\centering
\caption{Robustness against various post-processing on generated images.}
\setlength{\tabcolsep}{9.5pt}
\begin{tabular}{lcccc}
\hline
\textbf{Post-processing} & \textbf{BER}(\%)$\downarrow$ & \textbf{SR}(\%)$\uparrow$ & \textbf{T-A}$\uparrow$ & \textbf{I-A}$\uparrow$ \\ \hline
None                                               & 0.3 & 97.4 & 24.78 & 79.73 \\
Color Jitter                                       & 3.1 & 91.1 & 24.78 & 76.69 \\
Crop \& Resize                                     & 3.6 & 80.8 & 24.22 & 78.65 \\
Rotation                                           & 1.3 & 80.8 & 24.22 & 70.36 \\
Blur                                               & 1.5 & 90.6 & 24.53 & 79.42 \\
Gaussian Noise                                     & 5.3 & 78.2 & 24.76 & 78.48 \\
JPEG Compress                                      & 1.9 & 89.2 & 24.23 & 79.37 \\
Sharpness                                          & 0.4 & 96.9 & 24.57 & 79.48 \\ \hline
\end{tabular}
\label{tab: postprocessing distortion}
\end{table}

With generated images, the attacker may post-process them to remove the watermark.
We evaluated common post-processing methods: color jitter, crop \& resize, rotation, blur, adding Gaussian noise, JPEG compress, and sharpness.
Referring to Table \ref{tab: postprocessing distortion}, our method exhibits acceptable robustness in most cases. For some slightly higher BER such as Gaussian Noise, the corresponding utility of the processed images also degrades in terms of visual quality (\ie, T-A and T-I).
Some visual examples are provided in Appendix \ref{app:Post-processing distortion Visual}.
Detail settings of post-processings can also be found in Appendix \ref{app:Post-processing distortion details}.

\subsubsection{Robustness Against Pre-processing on Concept}
\label{sec:pre-p}
\begin{table}[t]
\centering
\caption{Robustness against different pre-processing on the watermarked concept. }
\setlength{\tabcolsep}{9.5pt}
\begin{tabular}{lcccc}
\hline
\textbf{Pre-processing} & \textbf{BER}(\%)$\downarrow$ & \textbf{SR}(\%)$\uparrow$ & \textbf{T-A}$\uparrow$ & \textbf{I-A}$\uparrow$ \\ \hline
None                    & 0.3        & 97.4                 & 24.78                   & 79.73                    \\
Gaussian Noise          & 0.8        & 94.8                 & 25.35                   & 77.76                    \\
Rescaling                & 16.2       & 26.6                 & 27.06                   & 78.42                    \\
Smoothing               & 13.7       & 34.1                 & 27.81                   & 77.78                    \\ \hline
\end{tabular}
\label{tab:pre-processing}
\end{table}

\vspace{-0.5em}

\begin{figure}[]
\includegraphics[width=0.42\textwidth]{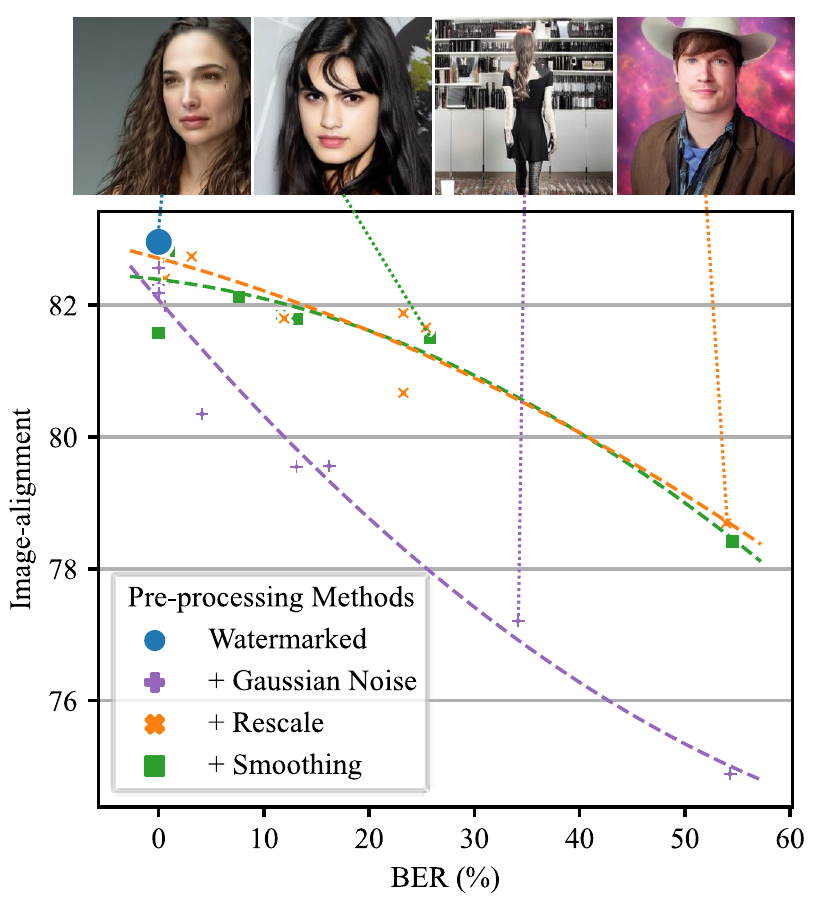}
\caption{Image-alignment and BER trade-off for concept pre-processing distortions. The blue dot represents the watermarked concept without pre-processing. Visual examples are presented at the top of the figure.}
\label{fig:diffip_preprocess tradoff}
\end{figure}

The attacker can also pre-process the download concept before generating images.
We consider the following three pre-processing operations:

\noindent \textit{\underline{Adding Gaussian Noise.}}
We added Gaussian noise to the embeddings with an intensity of $\sigma=1 \times 10^{-2}$. We examined 36 Textual Inversion embeddings downloaded from the internet, with the norm typically around $1\times 10^{-1}$. \Tref{tab:pre-processing} demonstrates our method's robustness against Gaussian noise with this intensity. It can be observed that our method demonstrates good performance against Gaussian noise of this intensity.

\noindent \textit{\underline{Rescaling Concept Embedding.}}
Another operation with relatively minor semantic impact is rescaling the embeddings, simply multiplying the embedding with a factor. In our experiments, we set this value to 0.25.  \Tref{tab: Robustness fine-tuning} shows that although the extraction rate of our method decreases to some extent with rescaling, the Image-Alignment (I-A) also decreases, indicating a strong correlation between the drop in watermark extraction rate and the decline in image fidelity caused by rescaling.

\noindent \textit{\underline{Smoothing Concept Embedding.}}
The third operation involves smoothing the embeddings through a conv layer with the 1D kernel [0.1, 0.2, 0.4, 0.2, 0.1]. Our experiments reveal that while the extraction rate of our method experiences a certain level of decrease, the Image-Alignment (I-A) also declines simultaneously.

Furthermore, we conducted more detailed testing on a single concept. In \Fref{fig:diffip_preprocess tradoff}, we present the Image-alignment and BER trade-off for different pre-processing distortions. This graph demonstrates that for these three pre-processing distortions, attackers attempting to reduce the extraction Success Rate must inevitably sacrifice a portion of image fidelity. Interestingly, attackers using Rescale and Smooth achieve better results compared to adding Gaussian Noise. This could be attributed to Gaussian Noise being isotropic, leading to a higher probability of introducing distortions orthogonal to the original embedding direction. As a result, the semantic deviation becomes more severe compared to Rescale and Smooth, causing the Image-alignment metric to decline more rapidly with increasing distortion.

\subsection{Adpative Attacks}

\label{sec:aa}

\begin{table}[]
\centering
\caption{The performance of our method against adaptive attacks (\ie, retrain attack and forgery attack). }
\setlength{\tabcolsep}{8pt}
\begin{tabular}{lcccc}
\hline
\textbf{Adaptive Attack}     & \textbf{BER(\%)$\downarrow$} & \textbf{SR(\%)$\uparrow$} & \textbf{T-A$\uparrow$} & \textbf{I-A$\uparrow$} \\ \hline
None                         & 0.0                    & 100                           & 25.39                  & 82.07                  \\
Retrain Embedding            & 0.9                    & 93.8                           & 21.05                  & 81.71                  \\
Forgery Attack               & 4.9                    & 67.2                           & 24.20                  & 81.73                  \\
\hline
\end{tabular}
\label{tab:adaptive attack}
\end{table}

In addition to the aforementioned watermark removal methods, there are several other ways attackers can attempt to remove watermarks once they have identified their presence. We primarily considered the following two approaches: retrain embedding and forgery attack. Due to the significant cost involved in executing these attacks, we conducted experiments on a single concept for each approach.

\subsubsection{Retrain Concept Embedding}
Attackers can retrain a new concept by employing the Textual Inversion pipeline on the images sampled from the obtained concept. We simulated this attack on one concept and found that our watermark demonstrated robust resistance to this attack. Referring to Table \ref{tab:adaptive attack}, even when attackers used the retrained concept to generate images, we still achieved a 99.1\% bit accuracy. Another observed phenomenon was the decline in the Text-alignment metric for the retrained concept, indicating a reduction in its editability.

\subsubsection{Forgery Attack}
Attackers can try to execute the watermarking pipeline on the acquired concept. As a matter of fact, the encoder is privately owned by the concept owner and the honest platform, making it practically inaccessible to potential attackers. What attackers can attempt is to obtain the embeddings and then utilize this pipeline to reinitialize an encoder and decoder, subsequently adding a watermark. They may try to use this new watermark to misguide the original decoder toward an incorrect bit string.

Referring to \Tref{tab:adaptive attack}, if the attackers train a new encoder and decoder from scratch using this pipeline, the new concept preserves the original semantics reasonably well, and we still achieve a good 95.1\% bit accuracy.

\afterpage{
\begin{table}
\centering
\caption{The influence of encoder architectures.}
\setlength{\tabcolsep}{9pt}
\begin{tabular}{l c c c c}
\hline
\textbf{Architectures} & \textbf{BER(\%)$\downarrow$} & \textbf{SR(\%)$\uparrow$} & \textbf{T-A$\uparrow$} & \textbf{I-A$\uparrow$} \\
\hline
\baseline U-Net & 0.3 & 98.0 & 25.03 & 81.76 \\
Noise LUT & 14.5 & 37.6 & 26.75 & 82.77 \\
MLP & 0.5 & 96.5 & 22.21 & 82.67 \\
Shallow MLP & 0.2 & 98.8 & 22.65 & 83.24 \\
Transformer & 26.0 & 9.0 & 22.79 & 84.06 \\
\hline
\end{tabular}
\label{tab:encoder_arch}
\vspace{-1em}
\end{table}

\begin{figure}[h!]
\centering
\includegraphics[width=0.40\textwidth]{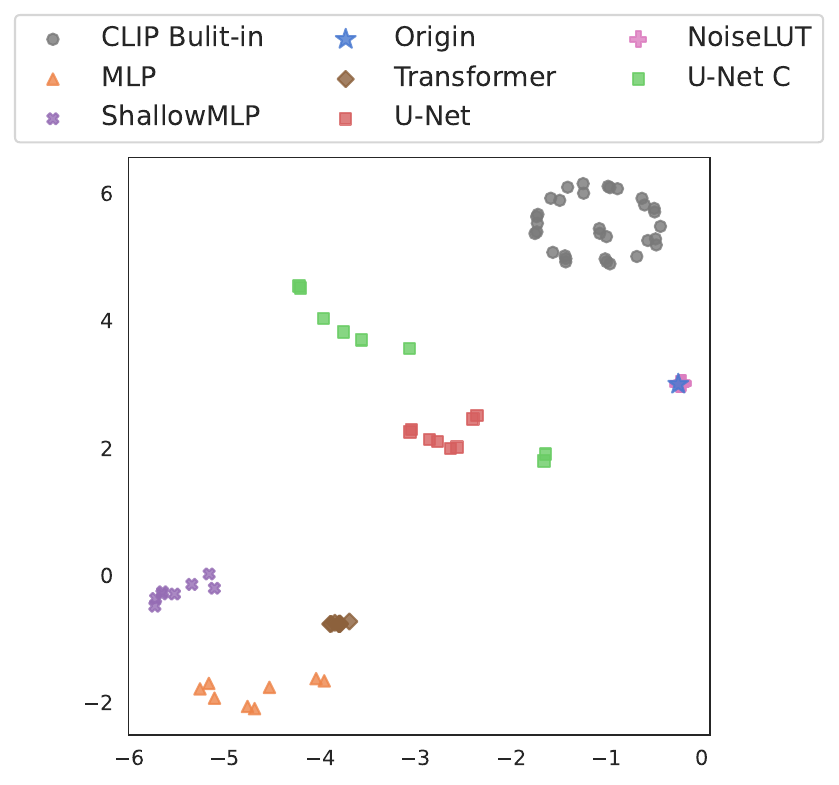}
\caption{T-SNE for watermarked embeddings under different encoder architectures. For an encoder, different points represent different watermarked messages. \textbf{Origin} represents the original embeddings. \textbf{CLIP Built-in} denotes some embeddings directly from the CLIP text encoder. \textbf{U-Net C} represents U-Net with contrastive loss applied during training. The rest of the figures show results from different architectures without using contrastive loss.}
\label{fig:t-SNE}
\end{figure}
}

\subsection{Ablation Studies}

We conducted comprehensive ablation studies on our pipeline, focusing on a single concept due to experimental costs. For all experiments except the robustness fine-tuning section, we did not perform the second-stage fine-tuning to avoid its influence.

\subsubsection{The Influence of Encoder Architecture} \label{ab:Different Encoder Architecture}
In addition to the default U-Net, we further explore other various options, including Noise LUT, MLP, shallower MLP, and transformer, which will be first introduced below.

\noindent \textit{\underline{Noise LUT.}}
In the Noise LUT approach, we add fixed perturbations to the Textual Inversion embeddings. Specifically, for different watermark bit strings, distinct noise is added to the TI embedding, and this noise is pre-determined. During training, only a single decoder is required to discriminate between different watermarks and extract the corresponding bit string information from the images. This method is akin to a Look-Up Table (LUT), hence we refer to it as Noise LUT. We attempt various choices of noise variance, and in this study, we report the best result with $\sigma=0.05$. Other experimental results can be found in Appendix \ref{app:More Noise LUT Results}.

\noindent \textit{\underline{MLP \& shallow-MLP.}}
For the implementation of the MLP encoder, we flatten the concept and then repeat the Message $m$ to match the size of the embedding. The concatenated data is then sent to the MLP with a hidden size of 1536. The default MLP consists of 3 layers of Linear transformations, and the shallower version has only 2 layers.

\noindent \textit{\underline{Transformer.}}
For the implementation of the Transformer encoder, we treat each bit in the message as a token and concatenate it with the original Textual Inversion concept as input to the transformer encoder layer. Under the standard setting, we use two layers of transformer layers. For the output, we use the token at the same position as the Textual Inversion embedding input as the output of the watermarked Textual Inversion embedding. Regarding parameter initialization, we made certain adjustments by initializing all LayerNorm to 0.1. This was found to effectively prevent output values from becoming too large.

From Table \ref{tab:encoder_arch} \footnote{In \Tref{tab:encoder_arch}~$\sim$~\Tref{tab:grad_steps}, the gray cell denotes the default setting.}, we can observe that Noise LUT achieves a relatively low extraction accuracy, rendering it unsuitable for practical use. The MLP base encoder achieves close to 100\% accuracy in extraction, but it suffers a significant decrease in T-A, indicating a substantial loss in editability, which is unacceptable. Although the Transformer encoder exhibits excellent fidelity, the extraction accuracy does not seem to improve further. Ultimately, the U-Net strikes the best balance among these three evaluation metrics.

We conducted t-SNE dimensionality reduction analysis for watermarked embeddings, as shown in \Fref{fig:t-SNE}. We randomly selected some words from the CLIP text encoder's embeddings, which are referred to as ``built-in'' embeddings, and their distribution space is often called the word vector space. It can be observed that the original Textual Inversion embedding, does not entirely fall into the distribution of the ``built-in'' embeddings. Due to the small magnitude of Gaussian Noise added, the NoiseLUT embeddings almost overlap with the initial Textual Inversion embeddings.

From \Fref{fig:t-SNE}, we can see that contrastive loss enables the embeddings with different watermarks to be more distant from each other, enhancing robustness against preprocessing distortions. Importantly, the embeddings generated by U-Net as the encoder are much close to the word vector space, unlike those by MLP, shallower MLP, and Transformer, which lack editability. Interestingly, this observation aligns with the findings in the Textual Inversion\cite{gal2022image}, where the authors suggested that embeddings closer to the word space tend to exhibit better editability. Our experimental results here are consistent with this theory.

\subsubsection{The Influence of Watermark Bits }

\begin{table}
\centering
\caption{The influence of watermark bits. }
\setlength{\tabcolsep}{8.5pt}
\begin{tabular}{c c c c c}
\hline
\textbf{Watermark Bits} & \textbf{BER(\%)$\downarrow$} & \textbf{SR(\%)$\uparrow$} & \textbf{T-A$\uparrow$} & \textbf{I-A$\uparrow$} \\
\hline
4 & 0.2 & 99.2 & 25.26 & 83.23 \\
\baseline8 & 0.3 & 98.0 & 25.03 & 81.76 \\
12 & 1.0 & 90.6 & 25.71 & 81.70 \\
\hline
\end{tabular}
\label{tab:bits}
\end{table}

We evaluated our pipeline's performance by embedding 4, 8, and 12 bits into Textual Inversion embeddings, examining the extraction ability, fidelity, and editability. \Tref{tab:bits} presents our results. As the number of embedded bits increased, we observed a slight decrease in the extraction accuracy and fidelity. However, our method consistently maintains editability even with a higher number of hidden bits.

\subsubsection{The Influence of Gradient Backward Steps}

\begin{table}[]
\centering
\caption{The influence of gradient backward steps. }
\setlength{\tabcolsep}{9pt}
\begin{tabular}{c c c c c}
\hline
\textbf{Gradient Steps} & \textbf{BER(\%)$\downarrow$} & \textbf{SR(\%)$\uparrow$} & \textbf{T-A$\uparrow$} & \textbf{I-A$\uparrow$} \\
\hline
1 & 24.0 & 18.3 & 26.37 & 82.98 \\
2 & 13.0 & 23.0 & 25.55 & 82.93 \\
\baseline3 & 0.3 & 98.0 & 25.03 & 81.76 \\
\hline
\end{tabular}
\label{tab:grad_steps}
\vspace{-1em}
\end{table}

We examined the impact of retaining different numbers of gradient backward steps on the results. For Gradient backward steps=0 and 1, we did not wait for the loss curve to converge before stopping since its loss curve descend very slowly. We trained all settings for 10,000 steps.

In Table \ref{tab:grad_steps}, we observe that retaining more gradient backward steps leads to better results. Our current GPU limitations prevent us from increasing the number of gradient backward steps while keeping the batch size constant, but we anticipate further room for improvement in the future.

\subsubsection{The Importance of Contrastive Loss} \label{Impact of Contrastive Loss}

We tested the effect of the Contrastive Loss on improving robustness against pre-processing distortion in one of the concepts. As shown in Table \ref{tab:Contrastive Loss}, adding this loss significantly improved the performance against pre-processing distortion.

\begin{table}[t]
\centering
\caption{The impact of the contrastive loss on the robustness against pre-processing distortion}
\setlength{\tabcolsep}{22pt}
\begin{tabular}{lcc}
\hline
\multirow{2}{*}{\textbf{Pre-processing}} & \multicolumn{2}{c}{\textbf{Contrastive Loss (BER(\%)$\downarrow$)}} \\ \cline{2-3}
                                         & \textbf{w/o}           & \textbf{w/}          \\ \hline
None                                     & 0.7                   & 0.3                \\
Gaussian Noise                           & 4.7                   & 0.5                \\
Rescale                                  & 13.7                  & 3.5                \\
Smoothing                                & 22.3                  & 4.3                \\ \hline
\end{tabular}
\label{tab:Contrastive Loss}
\end{table}

\begin{table}[t]
\centering
\caption{
The impact of the robustness fine-tuning on the robustness against post-processing distortion}
\setlength{\tabcolsep}{15pt}
\begin{tabular}{lcc}
\hline
\multirow{2}{*}{\textbf{Post-processing}} & \multicolumn{2}{c}{\textbf{Robustness Fine-tuning (BER(\%)$\downarrow$)}} \\ \cline{2-3}
                                          & \textbf{w/o}             & \textbf{w/}             \\ \hline
None                                      & 0.2                    & 0.0                   \\
Color Jitter                              & 5.9                    & 2.5                   \\
Crop \& Resize                            & 3.5                    & 2.2                   \\
Rotation                                  & 3.0                    & 1.0                   \\
Blur                                      & 1.0                    & 0.3                   \\
Gaussian Noise                            & 16.8                   & 3.5                   \\
JPEG Compress                             & 0.5                    & 0.1                   \\
Sharpness                                 & 0.3                    & 0.0                   \\ \hline
\end{tabular}
\label{tab: Robustness fine-tuning}
\vspace{-1em}
\end{table}

\subsubsection{The Impact of Robustness Fine-tuning} \label{Robustness fine-tuning}

We investigated the influence of robustness fine-tuning on the results. Table \ref{tab: Robustness fine-tuning}  demonstrates the BER results before and after robustness fine-tuning for different distortion types. Across almost all distortion types, Robustness fine-tuning enhances the decoder's capabilities.

\section{Discussion}

\subsection{The Capacity of the Proposed Method.} 
The empirical results demonstrate that our method achieves a remarkable 97.4\% Success Rate for the evaluated concepts using 8-bit settings. This renders it suitable for accommodating up to 255 users (except for the held-out ``00..0'' for clean images). It is important to recognize that many concepts generated through Textual Inversion are not intended to encode precise messages and only few tokens are generally utilized to convey semantic information. The limited tokens inherently limit the overall information capacity and simultaneously pose challenges to user scalability. Regarding this, the Success Rate could be significantly enhanced by allocating a greater number of tokens in the Textual Inversion training at a relatively low additional cost. Moreover, expanding the gradient steps $k$ and the encoder and decoder model sizes can potentially improve scalability..

\subsection{The Computational Cost.}
\vspace{-1em}
Although the weights of the diffusion model do not need to be modified, the training process for the hiding and revealing nets involves the computation of a diffusion sampling pipeline, which can be computationally expensive.
Some recent work has demonstrated new training techniques for Textual Inversion\cite{fei2023gradient}. Additionally, more novel rapid sampling techniques are yet to be explored\cite{zhao2023unipc}\cite{lu2022dpmpp}, which may potentially be more efficient.

\subsection{The Concept-specific Encoder and Decoder.}
\vspace{-1em}
 In this work, we don't consider a unified watermark encoder and decoder which can be used among different concepts and allow adding watermarks to a new concept without training. Because the use of a shared encoder may introduce new vulnerabilities, such as the possibility for an attacker to target the fixed and shared encoder or decoder to learn how to remove watermarks or forge others.

\section{Conclusion}

In this paper, we point out the demand for guarding concept sharing (\ie, Textual Inversion).
To address it, we propose the novel \textit{concept watermarking}, wherein we provide some insights on training strategies and architecture design. We conduct extensive experiments to justify its practicability in concept-sharing scenarios, in terms of fidelity, effectiveness, integrity, and robustness. Moreover, adaptive attacks and ablation studies are also considered.

While concentrating on the technical aspects, we also advocate for the co-evolution of legislation and technology. By doing this, we hope to promote the coexistence of copyright protection and information transmission, paving the way for a time when original ideas can flourish while still being safeguarded.

\vspace{10mm}
\small


\bibliographystyle{unsrt}
\bibliography{ref}

\section{Appendix}

\subsection{Prompt Generation} \label{app:Prompt generation}

Due to the existence of various types of concepts, a universal prompt is not applicable. Therefore, we created different prompt datasets for different types of concepts.
We generate our training and testing prompt using GPT-3.5 by telling it with the following instructions:

\textit{[name] represents a (coarse class), take the following prompt as a reference, and generate (number) more prompts with various scenes and descriptions:
(prompt example)}

Here, (coarse class) should be replaced according to the category of the current concept, for example, person, style, car, crystalskull, etc. And (number) should be replaced by the number of prompts you want to generate.
For concepts related to particular persons or objects, we chose (A photo of [name]) as our prompt example. For the style-like concepts, we decided (A painting, art by [name]) as our prompt example. For cars and other objects in paper, we use (A photo of a [name])

We show 8 examples of the generated prompts for each category here:

\noindent \textbf{person}

\noindent\textit{``a photo of [name] playing guitar in the forest'' \\
``a photo of [name] shaking hands with Joe Biden'' \\
``a photo of mysterious [name] witcher at night'' \\
``a photo of [name] riding a bicycle under Eiffel Tower'' \\
``[name] wearing formal clothes, wearing a tophat and holding a cane.'' \\
``[name] reading a book with a prism on its cover'' \\
``[name] standing on a street corner'' \\
``a portrait of [name] with a crown and wearing a yellow t-shirt that has a space shuttle drawn on it.'' \\
}

\noindent \textbf{style}

\noindent\textit{``panda mad scientist mixing sparkling chemicals, art by [name]'' \\
``a dolphin in an astronaut suit on saturn, art by [name]'' \\
``The Oriental Pearl in the style of [name]'' \\
``A house with red roof in the style of [name]'' \\
``a mouse sitting next to a computer mouse, art by [name]'' \\
``a tiny dragon landing on a knight's shield, art by [name]'' \\
``an F1 race car in a Manhattan street, art by [name]'' \\
``an elephant on the full moon, art by [name]'' \\
}

\noindent \textbf{object-car}

\noindent\textit{ ``a picture of a [name] caught in the middle of a heavy downpour''\\
``an image of a [name] surrounded by a crowd of admirers at a car meet''\\
``a snapshot of a [name] speeding across the finish line at a race''\\
``a photograph of a [name] being loaded onto a transport truck for delivery''\\
``a still of a [name] parked under the stars at a remote campsite''\\
``an illustration of a [name] being chased by a pack of exotic sports cars''\\
``a panoramic shot of a [name] driving through a dense forest''\\
``a close-up of a [name] with custom modifications and unique paintwork''\\
}

\noindent \textbf{object-crystalskull}

\noindent\textit{ ``a photo of a [name] submerged in a clear pool of water'' \\
``an action shot of a [name] being used in a ritual'' \\
``a picture of a [name] next to a flickering candle'' \\
``an image of a [name] accompanied by a mysterious figure'' \\
``a snapshot of a [name] resting on an ornate pillow'' \\
``a photograph of a [name] being admired by a curious crowd'' \\
``a still of a [name] placed in the center of a sacred circle'' \\
``an illustration of a [name] floating in the depths of space'' \\
}

\subsection{Post-processing Distortion Visual Examples} \label{app:Post-processing distortion Visual}

\begin{figure}[h]
\centering
\includegraphics[width=0.5\textwidth]{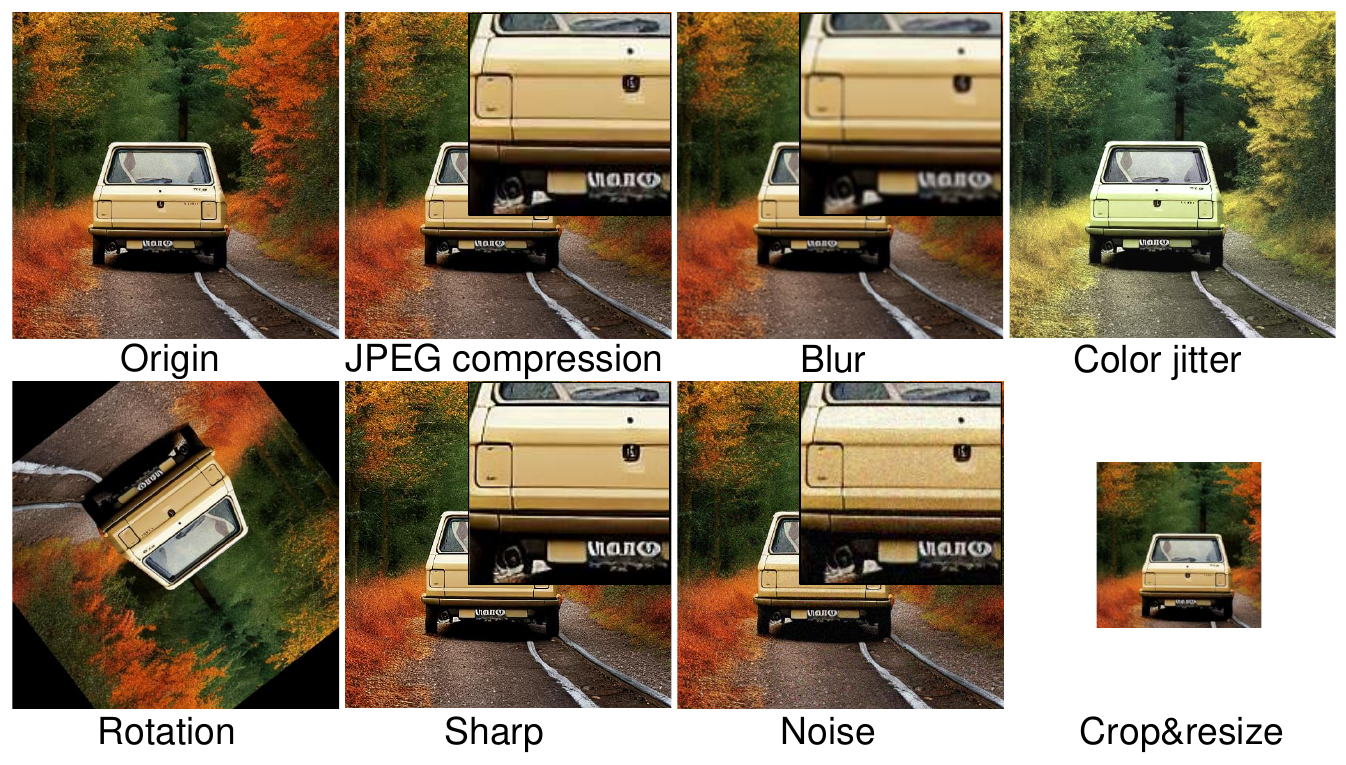}
\caption{Transformations evaluated in post-processing robustness.}
\label{fig:distortionshowcase}
\end{figure}

\Fref{fig:distortionshowcase} demonstrates the transformations we evaluated in the \Sref{Robustness}.

\begin{figure}[t]
\centering
\includegraphics[width=0.5\textwidth]{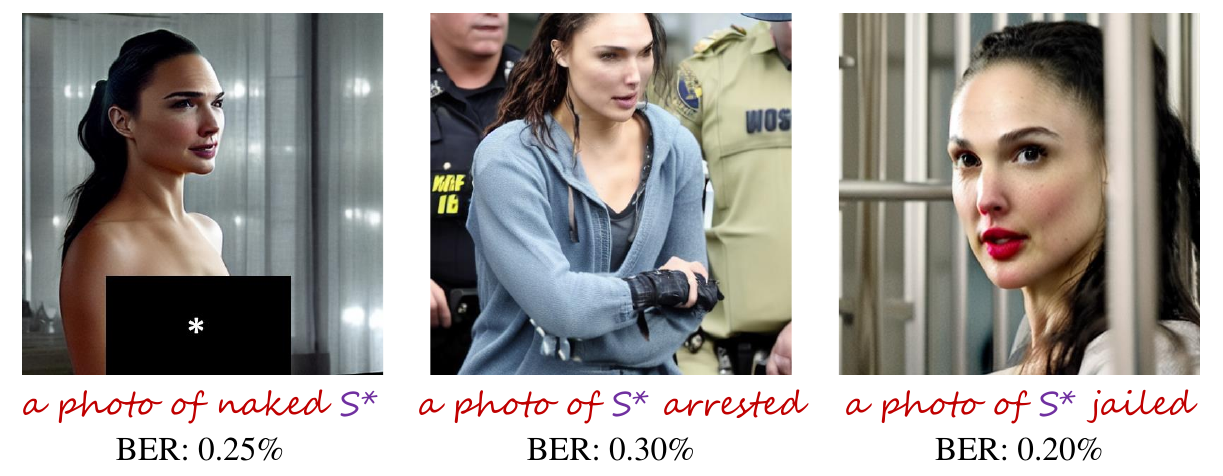}
\caption{Our method can be well performed on malicious prompts.}
\label{fig:misusecase}
\end{figure}

\subsection{Post-processing Distortion Details} \label{app:Post-processing distortion details}

We utilized the kornia python library for the following transformations: For Color jitter, we modified the brightness factor, contrast factor, saturation to 0.3, and hue factor to 0.1. For crop and resize, we randomly extracted $384 \times 384$ blocks from the $512 \times 512$ images and resized them to $256 \times 256$. Gaussian blur was applied with a kernel size of (3, 3) and sigma of (2.0, 2.0) on the images. Gaussian noise was added with a standard deviation of 0.05. JPEG compression was performed with a quality setting of 50. Rotation was randomly applied to the images within a range of 0 to 180 degrees. For sharpness, we set the intensity to 1.

All of the above distortions were applied after normalizing the images.

\subsection{Words Selected in t-SNE Plot}
Here we show the words we selected from CLIP text encoder built-in embedding in  \Fref{fig:t-SNE}:

 \textit{man  man$<$/w$>$  woman  woman$<$/w$>$  cat  cat$<$/w$>$  dog  \\ dog$<$/w$>$ car  car$<$/w$>$  water  water$<$/w$>$  happier$<$/w$>$ \\ happiest$<$/w$>$  happily$<$/w$>$  happiness$<$/w$>$  happy  happy$<$/w$>$ \\ sad  sad$<$/w$>$  fight  fight$<$/w$>$  fighter  fighter$<$/w$>$  fighters$<$/w$>$ \\ gal  gal$<$/w$>$  gado$<$/w$>$  trump  trump$<$/w$>$  person  person$<$/w$>$}

\subsection{More Noise LUT Results} \label{app:More Noise LUT Results}
Here, we present the results of the Noise LUT method under various standard deviation (std) conditions.

\begin{table}[h]
\centering
\begin{tabular}{ccccc}
\hline
\textbf{std}  & \textbf{BER}(\%)$\downarrow$ & \textbf{Success Rate}(\%)$\uparrow$ & \textbf{T-A}(\%)$\uparrow$ & \textbf{I-A}(\%)$\uparrow$ \\ \hline
$\sigma=0.01$ & 47.9        & 3.9                 & 25.99        & 81.41        \\
$\sigma=0.03$ & 26.9        & 11.8                 & 26.05        & 80.48        \\
$\sigma=0.05$ & 14.5        & 37.5                 & 26.57        & 80.55        \\
$\sigma=0.07$ & 17.6        & 29.3                 & 26.11        & 80.96        \\
$\sigma=0.08$ & 20.2        & 24.4                 & 26.99        & 79.59        \\
$\sigma=0.10$ & 27.9        & 13.0                 & 26.98        & 74.99        \\ \hline
\end{tabular}
\caption{Different results for Noise LUT method under different $\sigma$.}
\end{table}

When the sigma value is very small, the Image-alignment appears to be good, but the decoder struggles to distinguish between clean images and watermarked images. Conversely, when the sigma value is very large, the watermarked embedding does not resemble a typical embedding. In such cases, the embedding's impact on the generated results becomes unstable, leading to a decrease in both Image-alignment and Text-alignment.

\subsection{Against Malicious Prompt}

Our method is trained on general prompts and can effectively generalize across various types of prompts. Here in \Fref{fig:misusecase} we demonstrate that our approach achieves a high extraction rate for malicious prompts.

\subsection{Additional Visual Results}

\begin{figure*}
\centering
\includegraphics[width=1.0\textwidth]{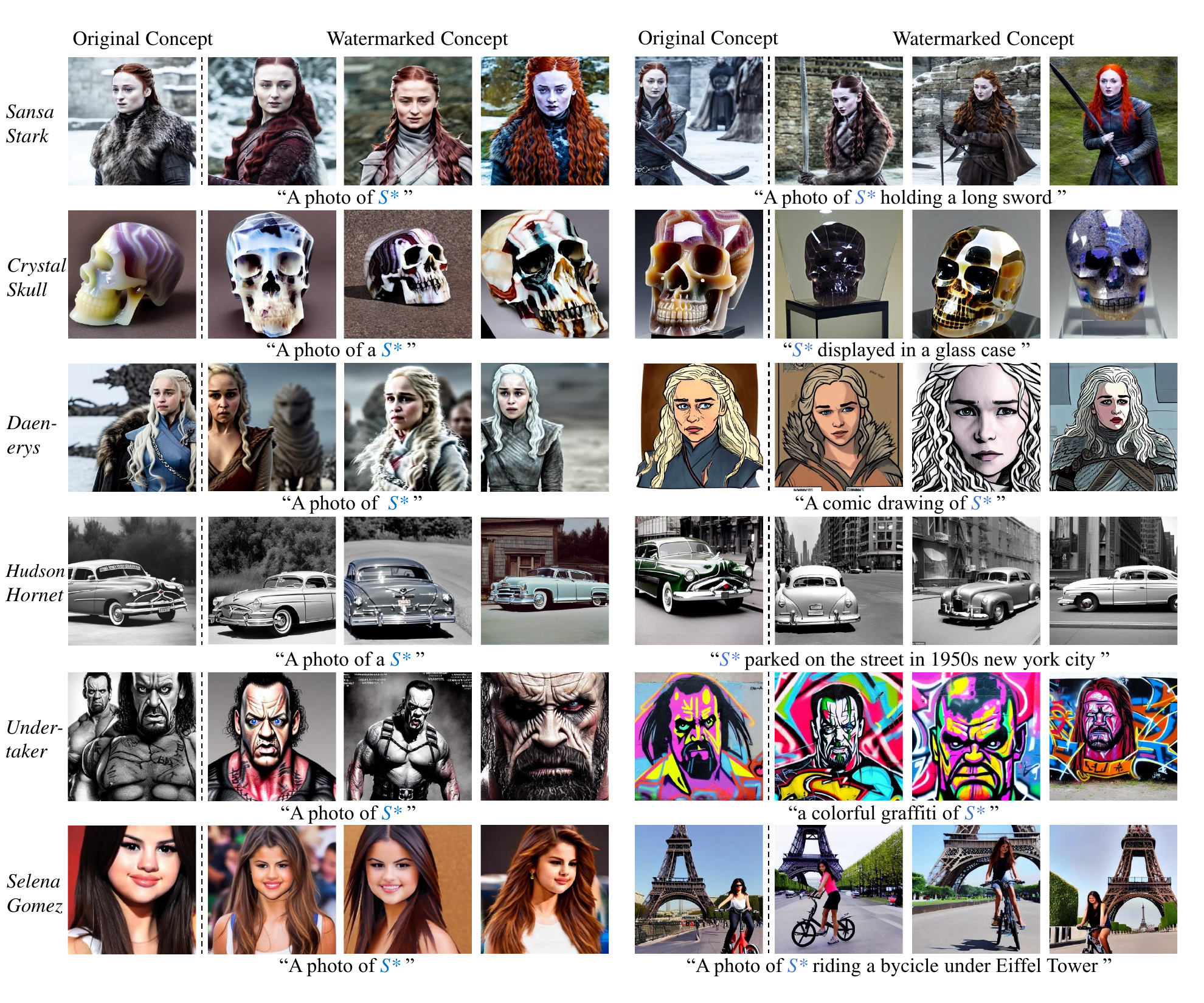}
\caption{Additional visual results.}
\label{fig:Additional visual results}
\end{figure*}

See \Fref{fig:Additional visual results}.

\subsection{More Discussion}
\label{app:d}

\noindent \textbf{The analysis of misidentification.} There exists a possibility of misidentifying a user (with watermarked message $\rvm_i$) as another user (with watermarked message $\rvm_j$, $j \neq i$). Suppose that the watermark decoder errs uniformly, and the assigned messages $\mathcal{M}$ are distributed randomly throughout the $q$-bit space $\{0,1\}^q$. In this case, the probability can be derived directly from the Success Rate (SR) as $P(\rvm \in \mathcal{M}-\{\rvm_i\}| \rvx^{\mathrm{wm}_i}) = (1-\mathrm{SR})\frac{|\mathcal{M}|}{2^q}$. This probability can be diminished by either increasing the bit number $q$ or reducing the assigned messages $|\mathcal{M}|$ while supposing the Success Rate is fixed.

\end{document}